\documentclass{article}

\usepackage{arxiv}

\usepackage{amssymb,amsmath}
\usepackage{float}
\usepackage{graphicx}
\usepackage{amsthm}
\usepackage{complexity}
\usepackage{color,soul}
\usepackage{hyperref}
\usepackage{subcaption}
\usepackage{algorithm}
\usepackage{diagbox}
\usepackage{multirow}
\usepackage{verbatim}
\usepackage[noend]{algpseudocode}

\graphicspath{{images/}}

\newcommand{\etal}{\textit{et al}.}
\newcommand{\round}[1]{\ensuremath{\lfloor#1\rceil}}
\newcommand{\neighbor}[1]{N_{\preccurlyeq #1}}
\newcommand{\priority}[1]{#1.\mathit{priority}}
\newcommand{\dd}[1]{#1.\mathit{decrease}}
\newcommand{\degree}[1]{#1.\mathit{degree}}
\newcommand{\visited}[1]{#1.\mathit{visited}}
\renewcommand{\G}{\mathcal{G}}
\DeclareMathOperator*{\argmax}{arg\,max}

\let\emptyset\varnothing

\let\epsilon\varepsilon

\algnewcommand\algorithmicforeach{\textbf{for each}}
\algdef{S}[FOR]{ForEach}[1]{\algorithmicforeach\ #1\ \algorithmicdo}

\title{High Quality Degree Based Heuristics for the Influence Maximization Problem}
\date{March 10, 2019}

\author{
	Maryam Adineh \\
	Department of Computer Engineering\\ Ferdowsi University of Mashhad\\ Mashhad, Iran \\
	\texttt{maryam.adineh@mail.um.ac.ir} \\
	\And
	Mostafa Nouri-Baygi\thanks{Corresponding author} \\
	Department of Computer Engineering\\ Ferdowsi University of Mashhad\\ Mashhad, Iran \\
	\texttt{nouribaygi@um.ac.ir} \\
}

\begin{document}
\maketitle

\begin{abstract}
The problem of influence maximization is to select the most influential individuals in a social network. With the popularity of social network sites, and the development of viral marketing, the importance of the problem has been increased. The influence maximization problem is $\NP$-hard, and therefore, there will not exist a polynomial-time algorithm to solve the problem unless $\P=\NP$. Many heuristics are proposed to find a nearly good solution in a shorter time.

In this paper, we propose two heuristic algorithms to find good solutions. The heuristics are based on two ideas: (1) vertices of high degree have more influence in the network, and (2) nearby vertices influence on almost analogous sets of vertices. We evaluate our algorithms on several well-known data sets and show that our heuristics achieve better results (up to $15\%$ in influence spread) for this problem in a shorter time (up to $85\%$ improvement in the running time).
\end{abstract}


\section{Introduction}
Interaction of people in a social network provides a lot of information about their behavior and the structure of the social graph. It has also made the social network a good platform to spread information, believes, innovation, and so on. One of the important applications of the spread of influence in social networks is viral marketing.

Consider a company that wants to market its product in a social network. A simple and low-cost approach is to select a subset of individuals to offer the product to them, so they will encourage their friends to buy it. This behavior is like spreading a virus in a society. The important portion of this type of marketing is the initial selection of the most influential individuals. This problem is known as the \emph{influence maximization} problem.

The influence maximization problem was first introduced by Domingos and Richardson~\cite{domingos2001mining,richardson2002mining}. Kempe \etal~\cite{kempe2003maximizing} formally defined the problem and proved that it is $\NP$-hard. They also introduced two monotone and submodular diffusion models for the spread of influence, namely independent cascade model and linear threshold model. An immediate result proved by Kempe \etal~\cite{kempe2003maximizing} was that a greedy hill climbing algorithm approximates the solution within $63\%$ of the optimal solution for these models.

Because the greedy algorithm runs a simulation several thousand times to find the marginal influence of each vertex, which is a time-consuming process, many heuristics are proposed to improve its performance. Although the heuristics have reduced the running time, they are still time-consuming for large-scale networks, which is the case for most social networks. On the other hand, degree based heuristics are very fast even on large-scale networks. Although they don't guarantee the quality of the solution, they still find good solutions for the problem.

In this paper, we propose two degree based heuristics with very short running time which improve the results of previous degree based heuristics. As will be illustrated by the experiments, the quality of the results produced by our algorithms are very close to the quality of the results produced by the greedy algorithm, while its running time is very small and close to the degree based heuristics.

This paper is an extended version of the paper presented before at the ICCKE' 2018 Conference~\cite{adineh2018maximum}. In this version a more detailed description of the algorithms are given and several new evaluations and comparisons are included.

The remaining of this paper is as follows. In Section~\ref{related work} the related work is reviewed. A formal definition of the problem is described in Section~\ref{preliminary}. We propose our heuristics in Section~\ref{proposed algorithm} and present the experimental results in Section~\ref{experiments}. Finally we conclude the paper in Section~\ref{conclusion}.

\section{Related Work} \label{related work}
The influence maximization problem was formally defined by Kempe \etal~\cite{kempe2003maximizing} and proved to be $\NP$-hard. They proposed a greedy hill climbing algorithm that yields a solution within $1-1/e-\epsilon$ factor of the optimal solution for two models they introduced for influence propagation. In the above approximation ratio, $e$ is the base of the natural logarithm, and $\epsilon$, which can be any positive real number, is the error of the Monte Carlo simulations. Picking a small value for $\epsilon$ increases the running time, while taking a large value for it reduces the quality of the result. In the algorithm of Kempe \etal, the most influential vertices are selected by their estimated marginal influence. Since estimated marginal influence is computed by a large number of simulations, the algorithm is not efficient.

In order to improve the efficiency of the computations, many studies have been made. Leskovec \etal~\cite{leskovec2007cost} proposed Cost-Effective Lazy Forward (CELF) optimization that reduces the computation cost of the influence spread using sub-modularity property of the objective function.

Chen \etal~\cite{chen2009efficient} proposed new greedy algorithms for independent cascade and weighted cascade models. They made the greedy algorithm faster by combining their algorithms with CELF. They also proposed a new heuristic, named \emph{degree discount}, which produces results of quality close to the greedy algorithm, while is much faster than that and performs better than the traditional degree and distance centrality heuristics.
 
In order to avoid running repeated influence propagation simulations, Borgs \etal~\cite{borgs2014maximizing} generated a random hypergraph according to reverse reachability probability of vertices in the original graph and select $k$ vertices that cover the largest number of vertices in the hypergraph. They guarantee $1-1/e-\epsilon$ approximation ratio of the solution with probability at least $1-1/n^l$. Later, Tang \etal~\cite{tang2014influence,tang2015influence} proposed TIM and IMM to cover drawbacks of Borgs \etal's algorithm~\cite{borgs2014maximizing} and improved its running time.
 
Bucur and Iacca~\cite{bucur2016influence} and Kr{\"o}mer and Nowakov{\'a}~\cite{kromer2017guided} used genetic algorithms for the influence maximization problem. Weskida and Michalski \cite{weskida2016evolutionary} used GPU acceleration in their genetic algorithm to improve its efficiency.

There are some community-based algorithms for the influence maximization problem that partition the graph into small subgraphs and select the most influential vertices from each subgraph. Chen \etal~\cite{chen2014cim} used H-Clustering algorithm and Manaskasemsak \etal~\cite{manaskasemsak2015community} used Markov clustering algorithm for community detection. Song \etal~\cite{song2015influence} divided the graph into communities, then selected the most influential vertices by a dynamic programming algorithm.

\section{Problem Definition}\label{preliminary}
In this section we formally define the influence maximization problem and the independent cascade diffusion model.

We consider a social network as an undirected graph $\G=(V,E)$ where $V$ is the set of individuals of size $n$, and $E$ is the set of relationships of size $m$. In this paper we describe the algorithms for undirected graphs, but it is trivial to extend the results to directed graphs. We also assume that $\G$ is unweighted, even though we can easily generalize the methods to the weighted case, where the weight of an edge $(u, v)$ denotes the probability of the influence propagation from $u$ to $v$. Clearly, the edge weights must be a value in the range $[0, 1]$

For each vertex $u$ and an integer $h > 0$, let $\neighbor{h}(u)$ denote the set of vertices of distance at most $h$ from $u$ in $\G$. We call $\neighbor{h}(u)$ the set of multi-hop neighbors of $u$.

For a set $S \subseteq V$ of vertices selected as the seed set to initiate the influence prorogation, let $I(S)$ denote the influence spread by $S$, i.e. the expected number of the influenced vertices, given $S$ as the initial seed set.

\subsection{Diffusion Model}
There are many diffusion models for the influence propagation process~\cite{kempe2015maximizing}. In this paper we focus on the \emph{Independent Cascade Model} (ICM). In the independent cascade model, for each edge $(u,v)$, a newly activated vertex $u$ can activate $v$ with probability $p_{u,v}\in [0,1]$.

The diffusion process is as follows. Let $S_i$ be the set of newly activated vertices in timestamp $i$. In timestamp $i+1$ each vertex $u\in S_i$ has a chance to activate each of its inactive neighbors. Once $u$ tried to activate its neighbor $v$, whether it succeeds or not, $u$ will not try to activate $v$ in later steps. Furthermore, each activated vertex remains active in all subsequent timestamps. This process terminates when no more activation is possible.

\subsection{Influence Maximization Problem}
In the problem of influence maximization, given a graph $\G$, a constant $k$ and a diffusion model $\mathcal{M}$, we are asked for a set $S$ of $k$ vertices with the maximum influence spread, $I(S)$. In this paper, we focus on the independent cascade model as $\mathcal{M}$, and leave extending the algorithms to other models to future work.

\section{Proposed Algorithms}\label{proposed algorithm}
In this section we describe our heuristics for the influence maximization problem under the independent cascade model.

As mentioned above, although the greedy algorithm and its variants guarantee the solution in terms of the influence spread, they are so time consuming especially for large scale social networks.
On the other hand, degree centrality heuristics don't guarantee the quality of the solution, but may produce solutions of high quality in much smaller time. As a result, we propose two novel heuristics based on degree centrality which demonstrate more influence spread in comparison to similar algorithms.

\begin{algorithm}
	\caption{MaximumDegree}\label{alg_maximum_degree}
	\begin{algorithmic}[1]
		\Require $\G(V,E)$: social network graph; $k$: size of the result set; $p$: propagation probability in the ICM.
		\Ensure $S$: seed set.

		\State $S\gets \emptyset$
		\For {$i \gets 1 \text{ to } k$}
		\State $u \gets \argmax_{v\in V \setminus S}d_v$ \Comment{Find the vertex $u$ with the highest degree}
		\State $S \gets S\cup\{u\}$
		\EndFor
		\State \Return {$S$}
	\end{algorithmic}
\end{algorithm}

Degree centrality heuristics select $k$ vertices with the highest degrees as the most influential vertices, because individuals with more relationships may have more influence spread in the network. The pseudocode of the maximum degree method is given in Algorithm~\ref{alg_maximum_degree}.

A variant of this method by Chen \etal~\cite{chen2009efficient}, called \emph{single discount}, decreases the degree of neighbors of each selected seed. For example when $u$ is selected as a seed, the degree of each neighbor $v$ is decreased according to the number of edges they have in common. Although these heuristics usually find suitable candidates as seeds, they are not good enough. The reason is that in social networks normally high degree vertices are close to each other and influence on almost similar sets of vertices.

To select better seed sets Chen \etal~\cite{chen2009efficient} proposed \emph{degree discount} which decreases degrees of vertices according to the expected number of adjacent active vertices and the amount of  influence propagation probabilities they have. Although the influence spread of degree discount is improved, it doesn't work well since it considers only direct neighbors and multi-hop influence spreads are not considered at all.

The main reason that degree centrality heuristics cannot keep up with greedy algorithms is that in social networks, vertices with high degrees are usually close to each other. Suppose that two adjacent vertices $u, v$ have the maximum degrees in the input graph. When we select $u$ as the first seed, with a high probability $v$ will also be activated by $u$. Therefore, there will not be much benefit from selecting $v$ as another seed. This will be amplified when the propagation probability, $p_{u, v}$, is higher. A similar case can be explained for multi-hop neighbors. In the following sections, we propose two heuristics to handle these situations properly.
 
\subsection{Removing Neighbors}
In the first heuristic, called \emph{NeighborsRemove}, we iteratively select $k$ vertices with the highest degrees. But to avoid selecting vertices with rather similar influence spread, in each step, we remove multi-hop neighbors of the selected seed from the list of candidates for subsequent steps.

More precisely, in the first iteration, we select the vertex $u$ with the maximum degree as the first seed. Since the multi-hop neighbors of $u$ will be directly influenced by $u$, even though they may have high degrees, we remove them from the list of candidates and select next seeds from the remaining vertices. This process terminates when $k$ seeds are selected.

In each step, when $u$ is selected as a seed, we remove its multi-hop neighbors at distance at most $h$, $\neighbor{h}(u)$, with a breadth-first search starting from $u$. An important parameter here is $h$, the maximum level at which the visited vertices in the breadth-first search is removed.

It is easy to see that when the distance between a seed vertex $u$ and another vertex $v$ increases, the probability that $v$ will be activated by $u$ decreases dramatically. This amount is equivalent to the product of the activation probabilities of the edges in the path from $u$ to $v$.

In our experiments, as most of the work in the literature, we assume the activation probability of each edge is constant, and equal to $p$. Based on this assumption, the value of $h$ is dependent only on $p$. According to our experiments on several data-sets, which are reported in Appendix~\ref{appendix}, the appropriate value for $h$ is computed by $\round{12 \sqrt{p}}$. The notation $\round{x}$ here means rounding $x$ to the nearest integer. The pseudocode of the method is given in Algorithm~\ref{alg_neighbors_remove}.

\begin{algorithm}
	\caption{NeighborsRemove}\label{alg_neighbors_remove}
	\begin{algorithmic}[1]
		\Require $\G(V,E)$: social network graph; $k$: size of the result set; $p$: propagation probability in the ICM.
		\Ensure $S$: seed set.
		
		\State $S \gets \emptyset$
		\State $V' \gets V$ \Comment{Make a copy of the vertices of $\G$.}
		\For{$i \gets 1 \text{ to } k$} 
		\State $u \gets \argmax_{v\in V'}d_v$
		\State $S \gets S\cup \{u\}$
		\State $V' \gets V' \setminus \neighbor{\round{12 \sqrt{p}}}(u)$ \Comment{Remove from $V'$ all nodes of distance}
		\Statex \Comment{at most $\round{12 \sqrt{p}}$ from $u$.}
		\EndFor
		\State \Return $S$
	\end{algorithmic}
\end{algorithm}

\subsection{Decreasing Degree}
The second heuristic for the influence maximization problem is called \emph{DegreeDecrease}.
Similar to the NeighborsRemove heuristic, the main idea here is to select vertices with the highest degrees. But to avoid selecting vertices with rather similar sets of influenced vertices, in each step, we reduce the priority of selecting vertices close the the selected seed. In each step, when a vertex $u$ is selected as the seed, the amount of the reduction in the priority of each vertex $v$ is calculated according to the number of different paths from $u$ to $v$, and their lengths. In the following, a more detailed description of the algorithm is given.
    
In the beginning, the priority of selecting each vertex $u$, denoted by $\priority{u}$ is equal to the degree of $u$. As the first seed, we therefore select the vertex $s_0$ with the maximum degree. Then for each vertex $v \in \neighbor{h}(s_0)$, we decrease the $\priority{v}$ to reduce the chance of $v$ being selected as subsequent seeds. In the second step, the vertex with the highest $\priority{u}$ is selected as the second seed. This process continues until $k$ vertices are selected as the seed set.

The probability of activating $v$ by $u$ is decreased as the length of the path from $u$ to $v$ increases. In addition, this probability increases as the number of paths from $u$ to $v$ increases. Therefore, the larger the number of paths or the smaller the path length from a vertex $u$ to a multi-hop neighbor $v$, the more reduction is applied on $\priority{v}$ when $u$ is selected as a seed. This is to reduce the chance of selecting vertices close to $u$ as subsequent seeds.

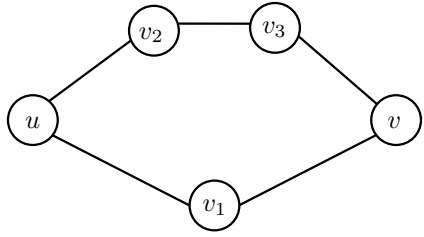
\begin{figure}
	\begin{center}
\begingroup%
  \makeatletter%
  \providecommand\color[2][]{%
    \errmessage{(Inkscape) Color is used for the text in Inkscape, but the package 'color.sty' is not loaded}%
    \renewcommand\color[2][]{}%
  }%
  \providecommand\transparent[1]{%
    \errmessage{(Inkscape) Transparency is used (non-zero) for the text in Inkscape, but the package 'transparent.sty' is not loaded}%
    \renewcommand\transparent[1]{}%
  }%
  \providecommand\rotatebox[2]{#2}%
  \newcommand*\fsize{\dimexpr\f@size pt\relax}%
  \newcommand*\lineheight[1]{\fontsize{\fsize}{#1\fsize}\selectfont}%
  \ifx\svgwidth\undefined%
    \setlength{\unitlength}{158.32725104bp}%
    \ifx\svgscale\undefined%
      \relax%
    \else%
      \setlength{\unitlength}{\unitlength * \real{\svgscale}}%
    \fi%
  \else%
    \setlength{\unitlength}{\svgwidth}%
  \fi%
  \global\let\svgwidth\undefined%
  \global\let\svgscale\undefined%
  \makeatother%
  \begin{picture}(1,0.54674533)%
    \lineheight{1}%
    \setlength\tabcolsep{0pt}%
    \put(0,0){\includegraphics[width=\unitlength,page=1]{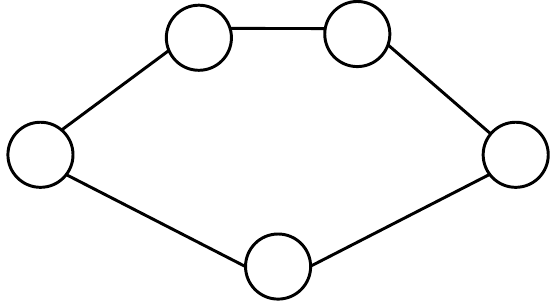}}%
    \put(0.05501856,0.24630398){\color[rgb]{0,0,0}\makebox(0,0)[lt]{\lineheight{1.25}\smash{\begin{tabular}[t]{l}$u$\end{tabular}}}}%
    \put(0.91767778,0.24639864){\color[rgb]{0,0,0}\makebox(0,0)[lt]{\lineheight{1.25}\smash{\begin{tabular}[t]{l}$v$\end{tabular}}}}%
    \put(0.32617597,0.45956479){\color[rgb]{0,0,0}\makebox(0,0)[lt]{\lineheight{1.25}\smash{\begin{tabular}[t]{l}$v_2$\end{tabular}}}}%
    \put(0.61923494,0.47156325){\color[rgb]{0,0,0}\makebox(0,0)[lt]{\lineheight{1.25}\smash{\begin{tabular}[t]{l}$v_3$\end{tabular}}}}%
    \put(0.47712422,0.04861474){\color[rgb]{0,0,0}\makebox(0,0)[lt]{\lineheight{1.25}\smash{\begin{tabular}[t]{l}$v_1$\end{tabular}}}}%
  \end{picture}%
\endgroup%

	\end{center}
	\caption{There are two different activation paths from $u$ to $v$.}
	\label{fig_different_paths}
\end{figure}

In the graph shown in Figure~\ref{fig_different_paths}, there are two different paths from $u$ to $v$. In each path, there is a possibility of $v$ being activated by $u$. The probability of activation of $v$ from the lower path $u \rightarrow v_1 \rightarrow v$ is greater than the upper path $u \rightarrow v_2 \rightarrow v_3 \rightarrow v$. In the lower path, $v$ will be activated when both edges $(u, v_1)$ and $(v_1, v)$ propagate the influence, which happens with probability $p^2$, while in the upper path, the probability is $p^3$, since three edges need to cooperate to propagate the influence.

Suppose when a vertex $u$ is selected as a seed, for each neighbor $v$ of $u$, we decrease $\priority{v}$ by a value that depends on $f(p)$, which is a function of $p$, the propagation probability of edges. We call $f(p)$ the \emph{path reduction coefficient}. In Figure~\ref{fig_different_paths}, $v$ may be activated by the lower path only if both edges $(u, v_1)$ and $(v_1, v)$ propagate the influence. Thus, we decrease $\priority{v}$ for this path by a value that depends on $f^2(p)$. Similarly for the upper path, we decrease $\priority{v}$ by a value that depends on $f^3(p)$, because the length of the path is 3.

The influence propagation through the two paths in Figure~\ref{fig_different_paths} are independent, so if we denote by $A$ (respectively $B$) the event of the influence propagation through the lower (resp. upper) path, the probability of the influence propagation through either of paths is equal to $$P(A \cup B) = P(A) + P(B) - P(A \cap B).$$ Events $A$ and $B$ are independents, so for the probability of the influence propagation through both paths we have $P(A \cap B) = p^5$. Given that $p < 1$ is a small value, $P(A \cap B)$ is negligible compared to $P(A)$ and $P(B)$. Therefore, to make computations simple, we can find the required reduction amount in $\priority{v}$ for each path independently, and then simply sum up those values.

\begin{figure}
	\begin{center}
\begingroup%
  \makeatletter%
  \providecommand\color[2][]{%
    \errmessage{(Inkscape) Color is used for the text in Inkscape, but the package 'color.sty' is not loaded}%
    \renewcommand\color[2][]{}%
  }%
  \providecommand\transparent[1]{%
    \errmessage{(Inkscape) Transparency is used (non-zero) for the text in Inkscape, but the package 'transparent.sty' is not loaded}%
    \renewcommand\transparent[1]{}%
  }%
  \providecommand\rotatebox[2]{#2}%
  \newcommand*\fsize{\dimexpr\f@size pt\relax}%
  \newcommand*\lineheight[1]{\fontsize{\fsize}{#1\fsize}\selectfont}%
  \ifx\svgwidth\undefined%
    \setlength{\unitlength}{212.10267795bp}%
    \ifx\svgscale\undefined%
      \relax%
    \else%
      \setlength{\unitlength}{\unitlength * \real{\svgscale}}%
    \fi%
  \else%
    \setlength{\unitlength}{\svgwidth}%
  \fi%
  \global\let\svgwidth\undefined%
  \global\let\svgscale\undefined%
  \makeatother%
  \begin{picture}(1,0.14358456)%
    \lineheight{1}%
    \setlength\tabcolsep{0pt}%
    \put(0,0){\includegraphics[width=\unitlength,page=1]{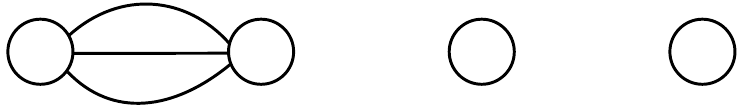}}%
    \put(0.04106943,0.05670677){\color[rgb]{0,0,0}\makebox(0,0)[lt]{\lineheight{1.25}\smash{\begin{tabular}[t]{l}$u$\end{tabular}}}}%
    \put(0.94047903,0.05670677){\color[rgb]{0,0,0}\makebox(0,0)[lt]{\lineheight{1.25}\smash{\begin{tabular}[t]{l}$v$\end{tabular}}}}%
    \put(0.63289024,0.05670677){\color[rgb]{0,0,0}\makebox(0,0)[lt]{\lineheight{1.25}\smash{\begin{tabular}[t]{l}$v_2$\end{tabular}}}}%
    \put(0.33122489,0.05670677){\color[rgb]{0,0,0}\makebox(0,0)[lt]{\lineheight{1.25}\smash{\begin{tabular}[t]{l}$v_1$\end{tabular}}}}%
    \put(0,0){\includegraphics[width=\unitlength,page=2]{multiple_paths.pdf}}%
  \end{picture}%
\endgroup%

	\end{center}
	\caption{There are $3 \cdot 2 \cdot 3$ different paths from $u$ to $v$.}
	\label{fig_multiple_paths}
\end{figure}

In Figure~\ref{fig_multiple_paths} there are $3 \cdot 2 \cdot 3$ different paths from $u$ to $v$. The length of each path is 3, and based on the above arguments, each path reduce a value from $\priority{v}$ that depends on $f^3(p)$, which totally sum up to $3 \cdot 2 \cdot 3 \cdot f^3(p)$. For ease of processing, we introduce a recurrence relation. Let $\dd{v}$ denote the value to be reduced from $\priority{v}$ for vertex $v$. The value of $\dd{v}$ for vertex $v$ in Figure~\ref{fig_multiple_paths} can be written as $$\dd{v} = \dd{v_2} \cdot c(v_2, v) \cdot f(p).$$ In the above recurrence relation, $c(u, v)$ denotes the number of edges in $\G$ from $u$ to $v$. The intuition is that in $\dd{v_2}$ we take into account both the number of different paths from $u$ to $v_2$ and the path reduction coefficient for those paths, $f^2(p)$. Therefore, it is enough to multiply $\dd{v_2}$ to the number of edges from $v_2$ to $v$ and the path reduction coefficient to determine $\dd{v}$.

We need to characterize $f(p)$ and the base case of the recurrence relation, to be able to update $\priority{v}$ for each vertex $v$. We choose two constant values $\alpha$ and $\beta$, whose exact value will be determined by further experiments, and define the functions based on these values. For the function $f(\cdot)$, we opt a linear function as $f(p) = \beta \cdot p$, and for the base case of the recurrence relation, we write $\dd{u} = \alpha$, where $u$ is the current seed vertex.

Since the value of $\dd{v}$ reduces as the distance of $u$ to $v$ grows, after enough hops, $\dd{v}$ gets close to 0 and we can stop further reduction process from $\priority{v}$. Based on our experiments, which are reported in Appendix~\ref{appendix}, we selected $\epsilon = 0.1$ as the threshold value for priority reduction. When the value of $\dd{v}$ falls below $\epsilon$, we stop further priority reduction propagation through $v$. If we select a small value as the threshold, the number of levels at which the breath-first search is performed is increased, and so is the running time. On the other hand, choosing a large value, reduces the number of levels of the breadth-first search, degrades the algorithm to normal maximum degree heuristic and decreases the accuracy.

{
\linespread{1.2}
\begin{algorithm}
	\caption{DegreeDecrease}\label{alg_degree_decrease}
	\begin{algorithmic}[1]
		\Require $\G(V,E)$: social network graph; $k$: size of the result set; $p$: propagation probability in the ICM.
		\Ensure $S$: seed set.
		
		\State $S \gets \emptyset$
		\State $V' \gets V$ \Comment{Make a copy of the vertices of $\G$.}
		\ForEach {$v \in V'$}
		\State $\priority{v} \gets \degree{v}$ \Comment{Initially, the priority of each vertex}
		\Statex \Comment{is equal to its degree.}
		\EndFor
		\For {$i \gets 1 \text{ to } k$}
		\State $u \gets \argmax_{v\in V'} \priority{v}$ \Comment{Select the vertex with}
		\Statex \Comment{the highest priority.}
		\State $S \gets S\cup\{u\}$
		\State $V' \gets V' \setminus \{u\}$
		\State Set $Q$ as an empty queue.
		\ForEach {$v \in V'$}
		\State $\visited{v}$ $\gets$ \textbf{false}
		\EndFor
		\State $\visited{u}$ $\gets$ \textbf{true} \Comment{Perform a BFS starting from $u$.}
		\State add $u$ to $Q$
		\State $\dd{u} \gets \alpha$
		\While {$Q$ is not empty}
		\State $v \gets$ extract the vertex with the minimum priority from $Q$
		\If {$\dd{v}>\epsilon$}
		\ForEach {$w \in v.\mathit{adj} \cap V'$ \textbf{and not} $\visited{w}$}
		\State $\visited{w}$ $\gets$ \textbf{true}
		\State add $w$ to $Q$
		\State $\dd{w} \gets \dd{v} \cdot c(v,w) \cdot \beta \cdot p$ \Comment{Compute the}
		\Statex \Comment{amount to be decreased from the priority of $w$.}

		\State $\priority{w} \gets \priority{w} - \dd{w}$
		\EndFor
		\EndIf
		\EndWhile
		\EndFor
		\State \Return $S$
	\end{algorithmic}
\end{algorithm}
}

Since the vertices with high influence in social networks usually have high degrees, choosing $\epsilon = 0.1$ resulted in both high accuracy and low running time. In addition, based on the experiments which are reported in Appendix~\ref{appendix}, the suitable value selected for $\alpha$ and $\beta$ are 50 and 10, respectively. The pseudocode of the method is given in Algorithm~\ref{alg_degree_decrease}.

\section{Experiments}\label{experiments}
In this section, we report and analyze the results of the experiments performed on the proposed heuristic algorithms and some previous algorithms using several real-life data-sets to evaluate the effectiveness of the new methods. We show that our maximum degree heuristics outperform previous degree based heuristics in terms of the spread of the influence, while output a solution of quality close to the approximation algorithms.

\subsection{Experimental Settings}
We evaluate our implementation on three data-sets which are commonly used in related researches, including~\cite{chen2009efficient}. The first data-set is NetHEPT with $n=15233$ and $m=58891$, the second data-set is NetPHY\footnote{Both of NetHEPT and NetPHY can be downloaded from \url{https://www.microsoft.com/en-us/research/wp-content/uploads/2016/02/weic-graphdata.zip}} with $n=37154$ and $m=231584$. These two networks are collaboration graphs crawled from \url{https://arxiv.org}, High Energy Physics -- Theory section and Physics section, respectively. The third data-set is Epinions~\cite{snapnets} which is a who-trust-whom online social network of a general consumer review site \url{http://epinions.com} with $n=75879$ and $m=508837$.

We compare our algorithms represented by NeighborsRemove and DegreeDecrease with four algorithms named SingleDiscount~\cite{chen2009efficient}, DegreeDiscount~\cite{chen2009efficient}, TIM~\cite{tang2014influence} and IMM~\cite{tang2015influence} that are available by their authors. All algorithms are implemented in C++ and compiled with GCC 6.2.1 and are run on a system with an Intel Core i7--3820 @ 3.60GHz and 32GB memory.
 
\subsection{Running Times and Influence Spread Analysis}
\begin{figure}
	\centering
	\begin{subfigure}[h]{0.49\textwidth}
		\fbox{\includegraphics[width=0.95\linewidth]{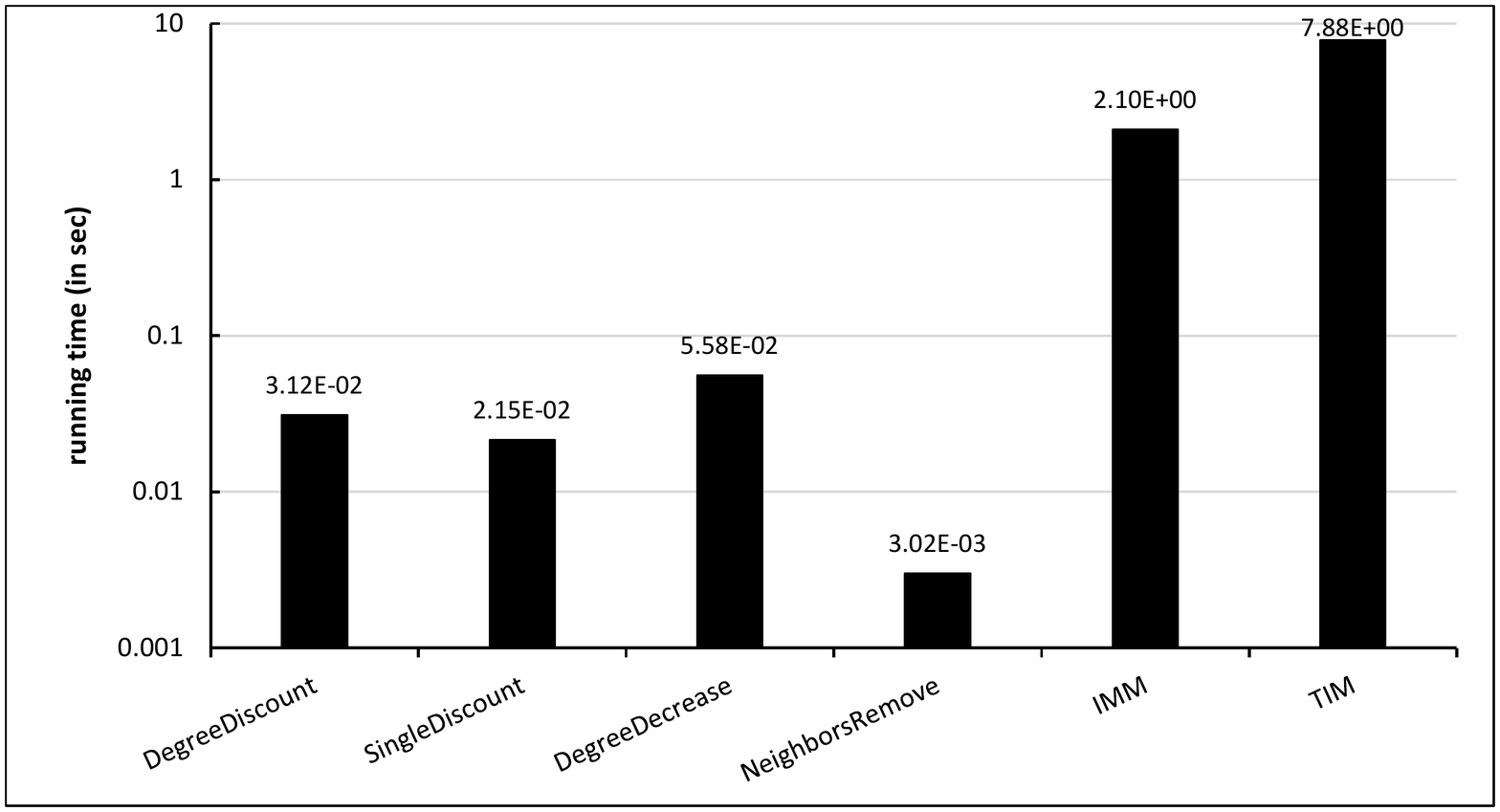}}
		\subcaption{}
		\label{runtime_hep_p_1percent}
	\end{subfigure}
	\begin{subfigure}[h]{0.49\textwidth}
		\fbox{\includegraphics[width=.94\linewidth]{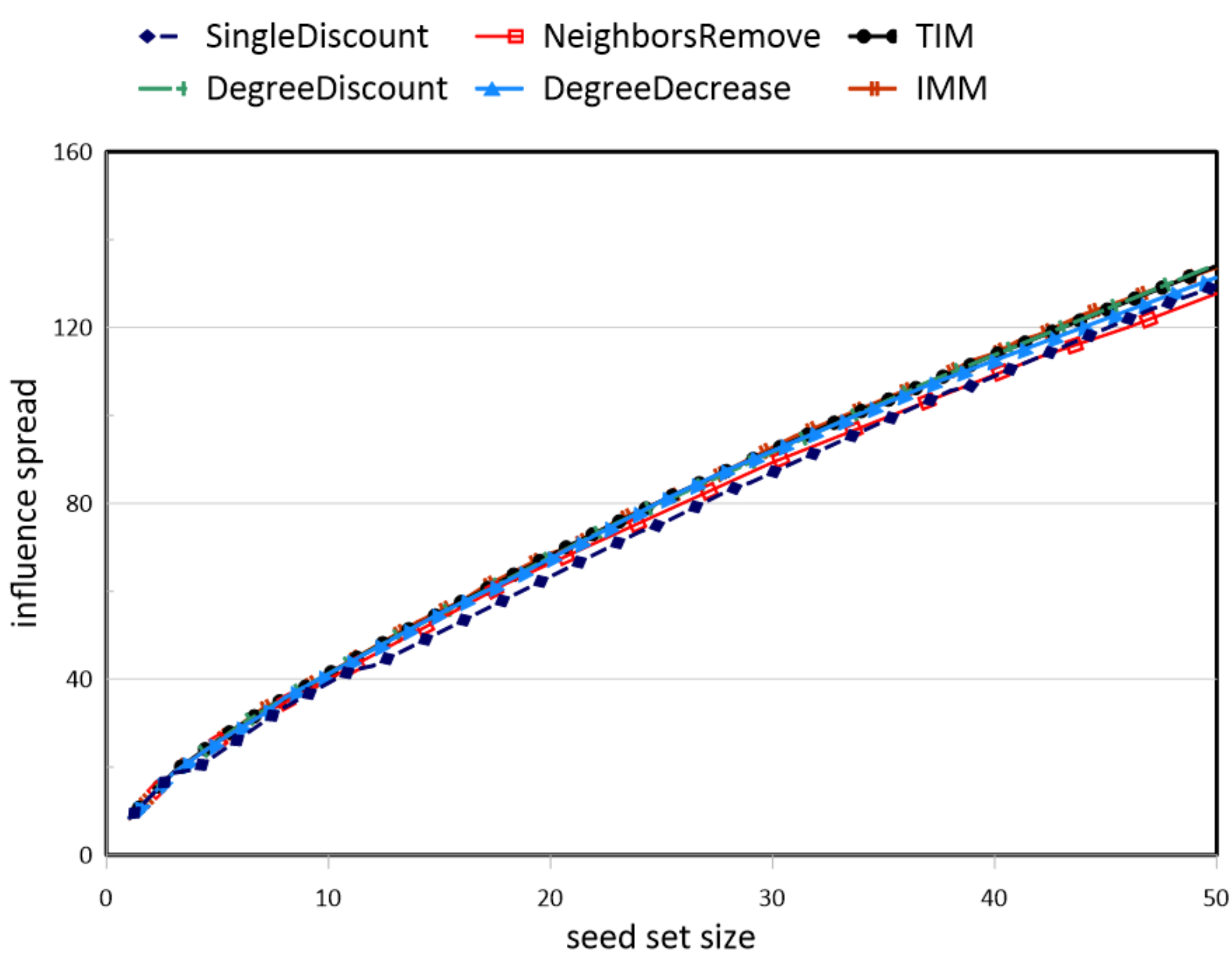}}
		\subcaption{}
		\label{inf_hep_p_1percent}
	\end{subfigure}
	\caption{Running times (a) and influence spreads (b) of algorithms on NetHEPT under independent cascade model ($p=0.01$, $k=50$).}
	\label{hep_p_1percent}
\end{figure}

\begin{figure}
	\centering
	\begin{subfigure}[h]{0.49\textwidth}
		\fbox{\includegraphics[width=0.95\linewidth]{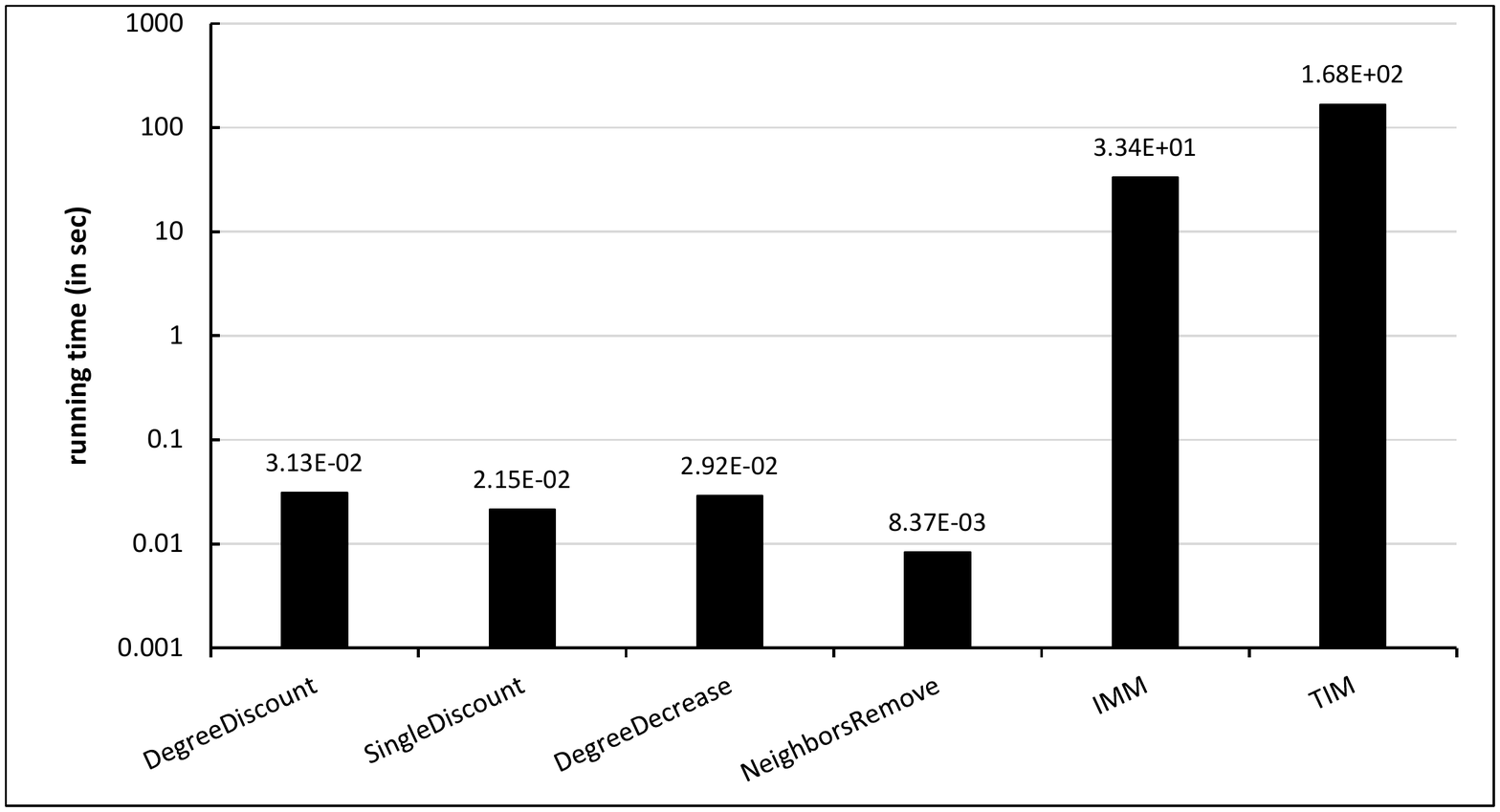}}
		\subcaption{}
		\label{runtime_hep_p_10percent}
	\end{subfigure}
	\begin{subfigure}[h]{0.49\textwidth}
		\fbox{\includegraphics[width=.95\linewidth]{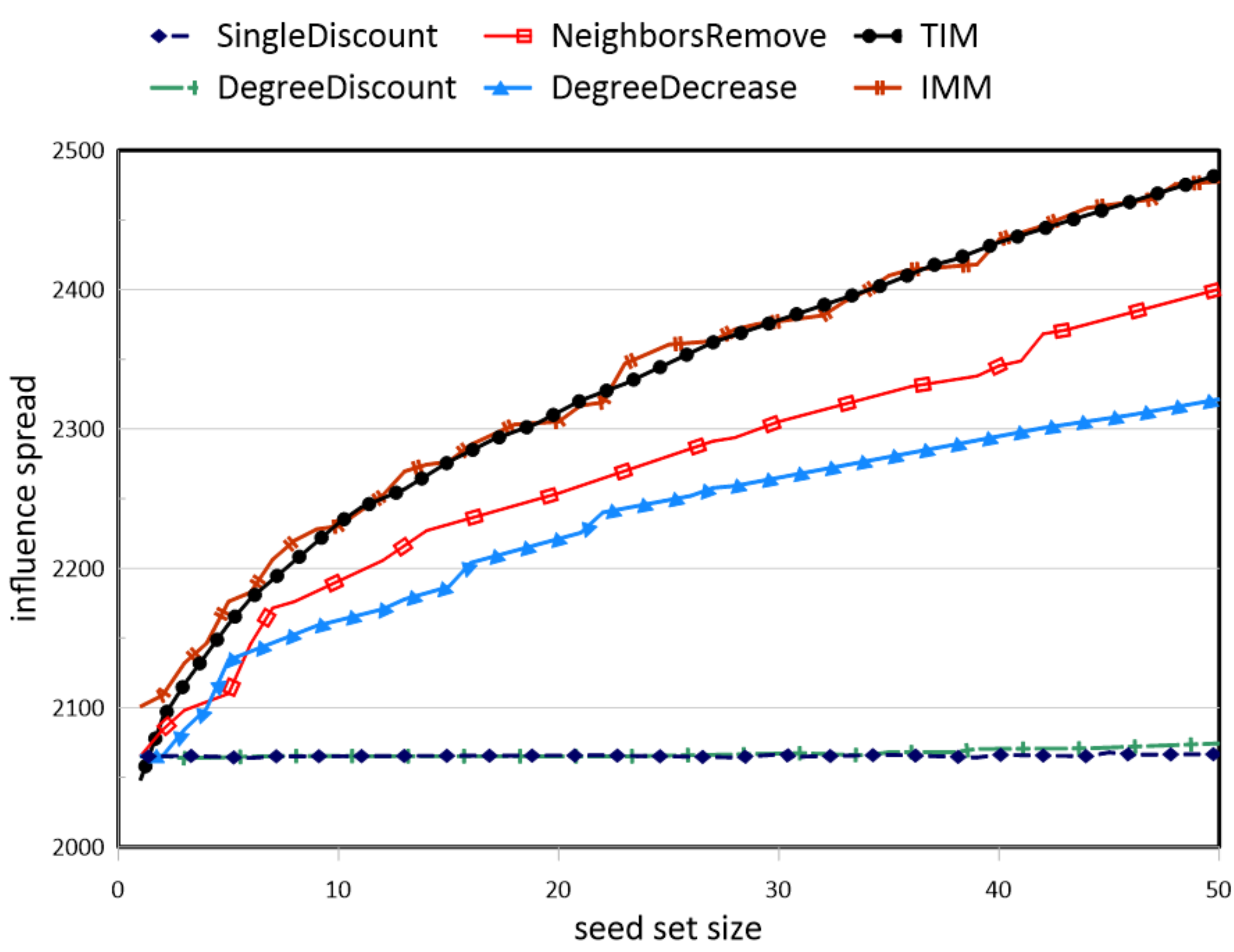}}
		\subcaption{}
		\label{inf_hep_p_10percent}
	\end{subfigure}
	\caption{Running times (a) and influence spreads (b) of algorithms on NetHEPT under independent cascade model ($p=0.1$, $k=50$).}
	\label{hep_p_10percent}
\end{figure}

\begin{figure}
	\centering
	\begin{subfigure}[h]{0.49\textwidth}
		\fbox{\includegraphics[width=0.95\linewidth]{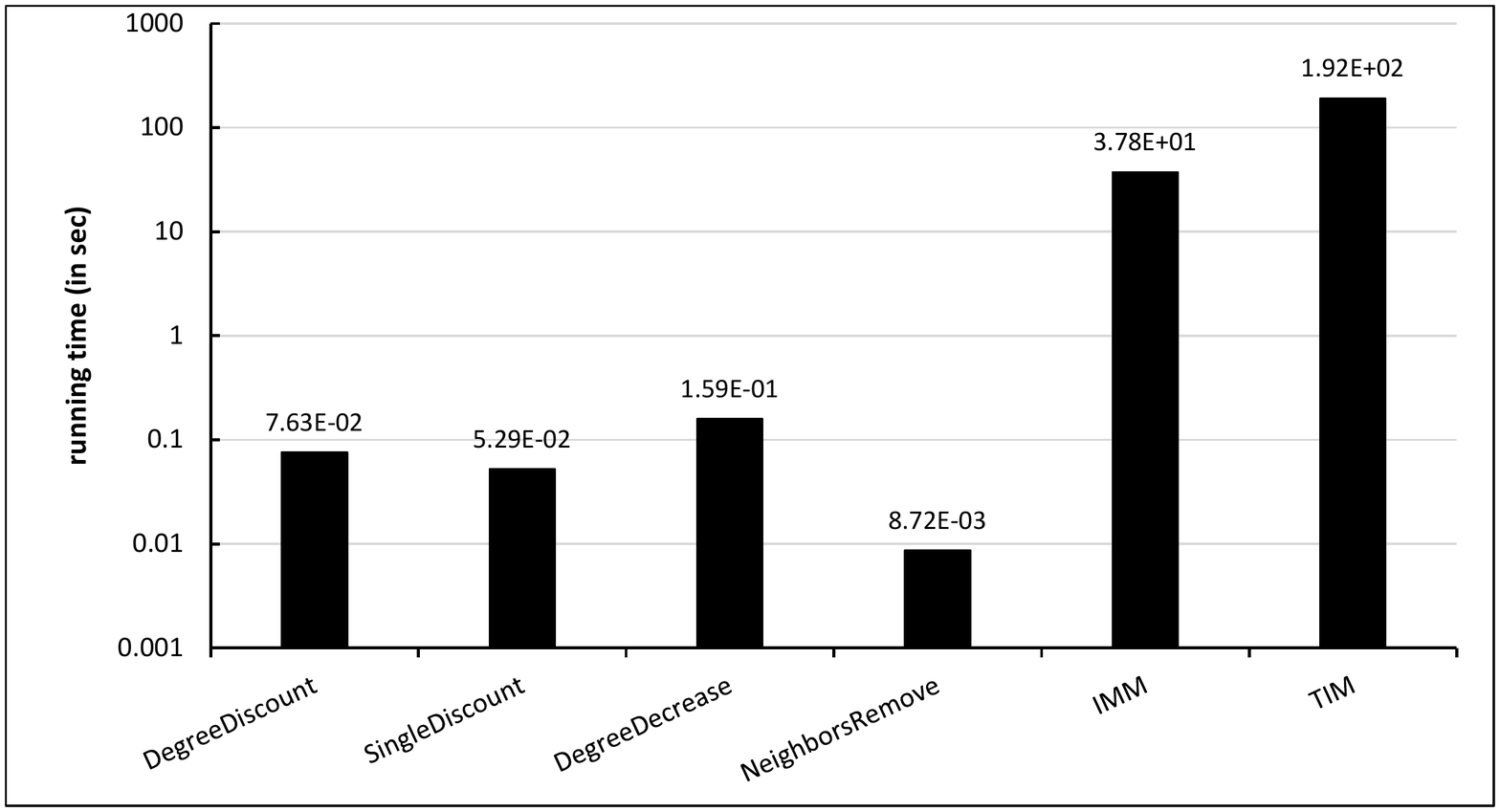}}
		\subcaption{}
		\label{runtime_phy_p_1percent}
	\end{subfigure}
	\begin{subfigure}[h]{0.49\textwidth}
		\fbox{\includegraphics[width=.94\linewidth]{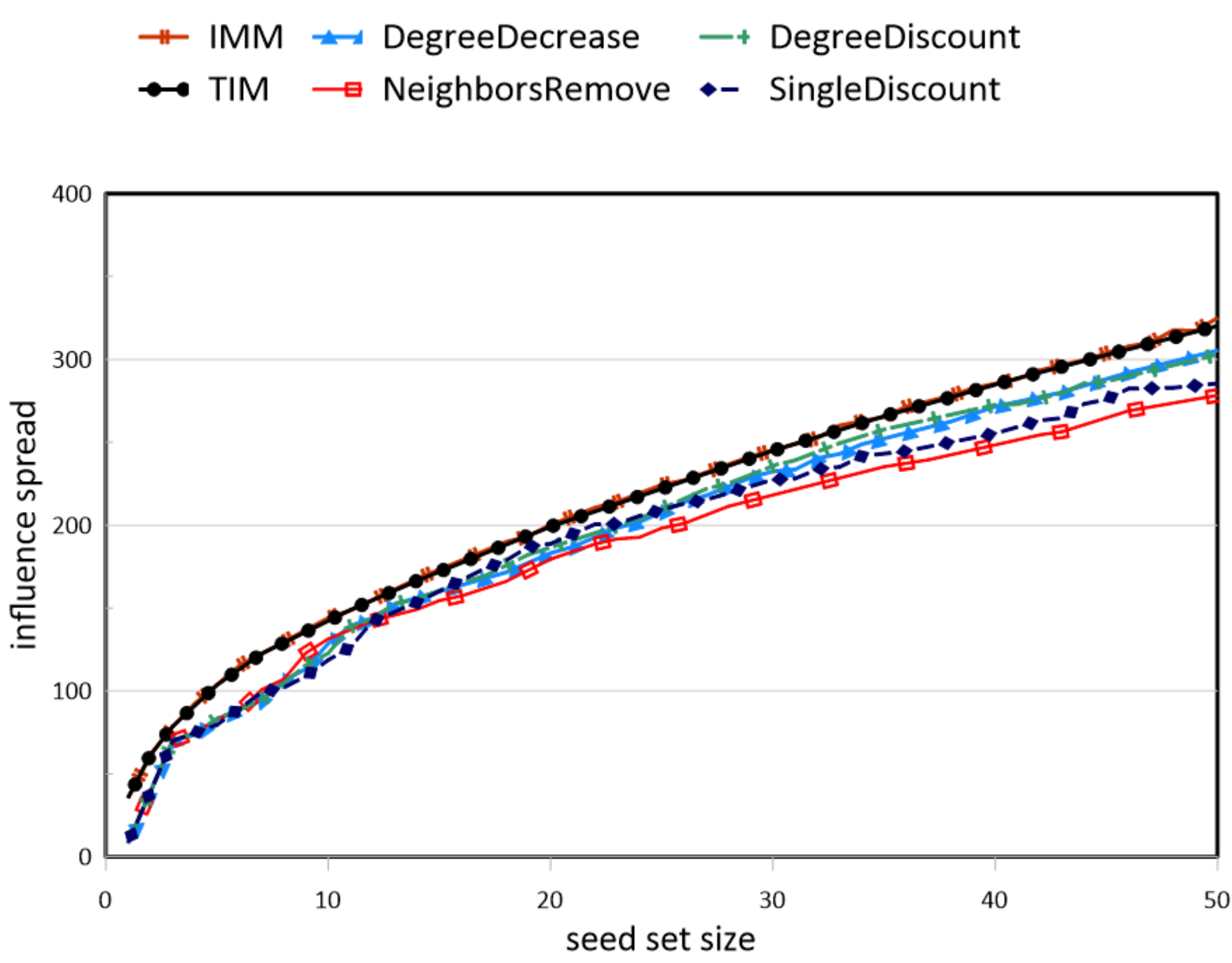}}
		\subcaption{}
		\label{inf_phy_p_1percent}
	\end{subfigure}
	\caption{Running times (a) and influence spreads (b) of algorithms on NetPHY under independent cascade model ($p=0.01$, $k=50$).}
	\label{phy_p_1percent}
\end{figure}

\begin{figure}
	\centering
	\begin{subfigure}[h]{0.49\textwidth}
		\fbox{\includegraphics[width=0.95\linewidth]{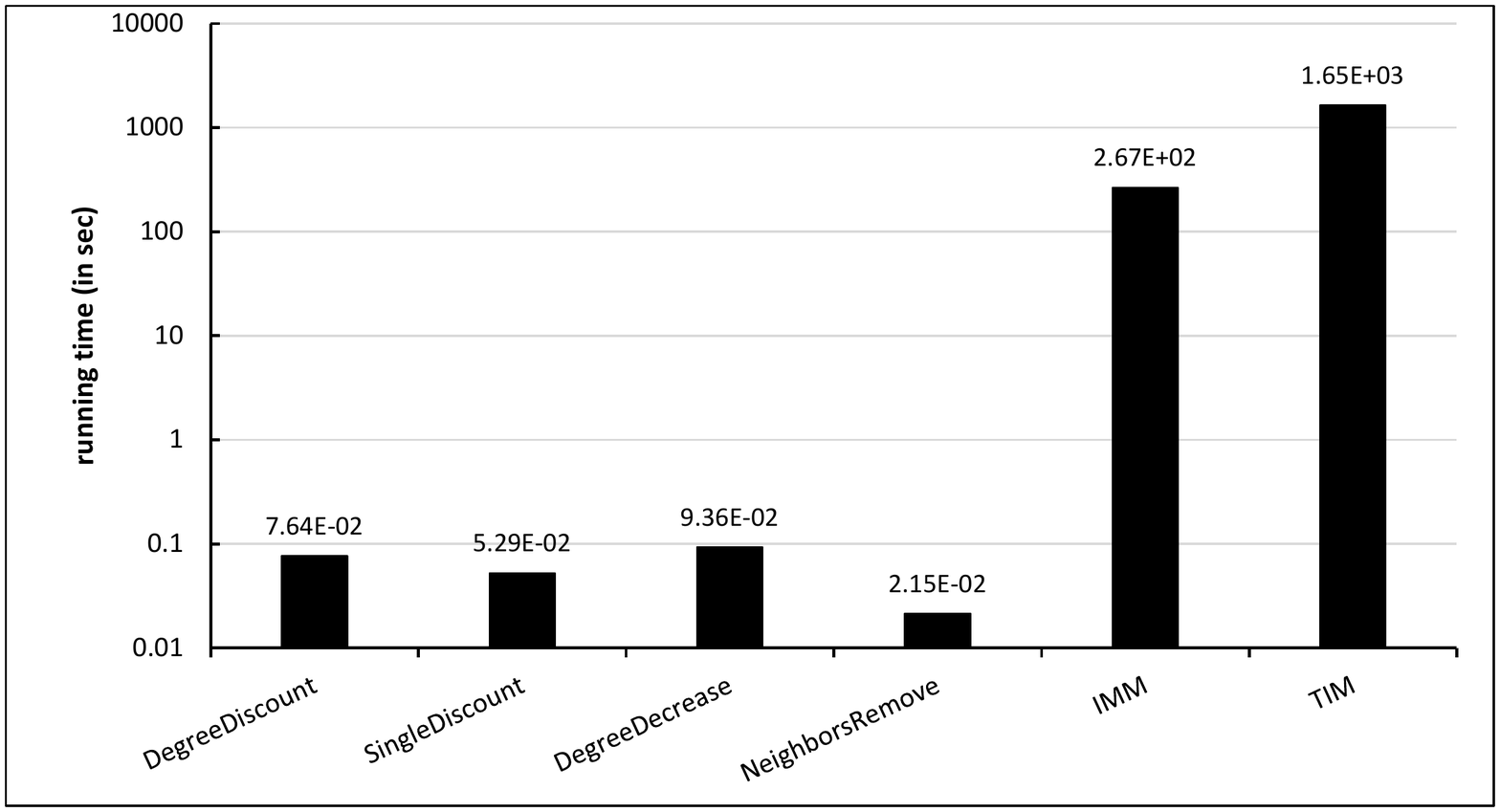}}
		\subcaption{}
		\label{runtime_phy_p_10percent}
	\end{subfigure}
	\begin{subfigure}[h]{0.49\textwidth}
		\fbox{\includegraphics[width=.95\linewidth]{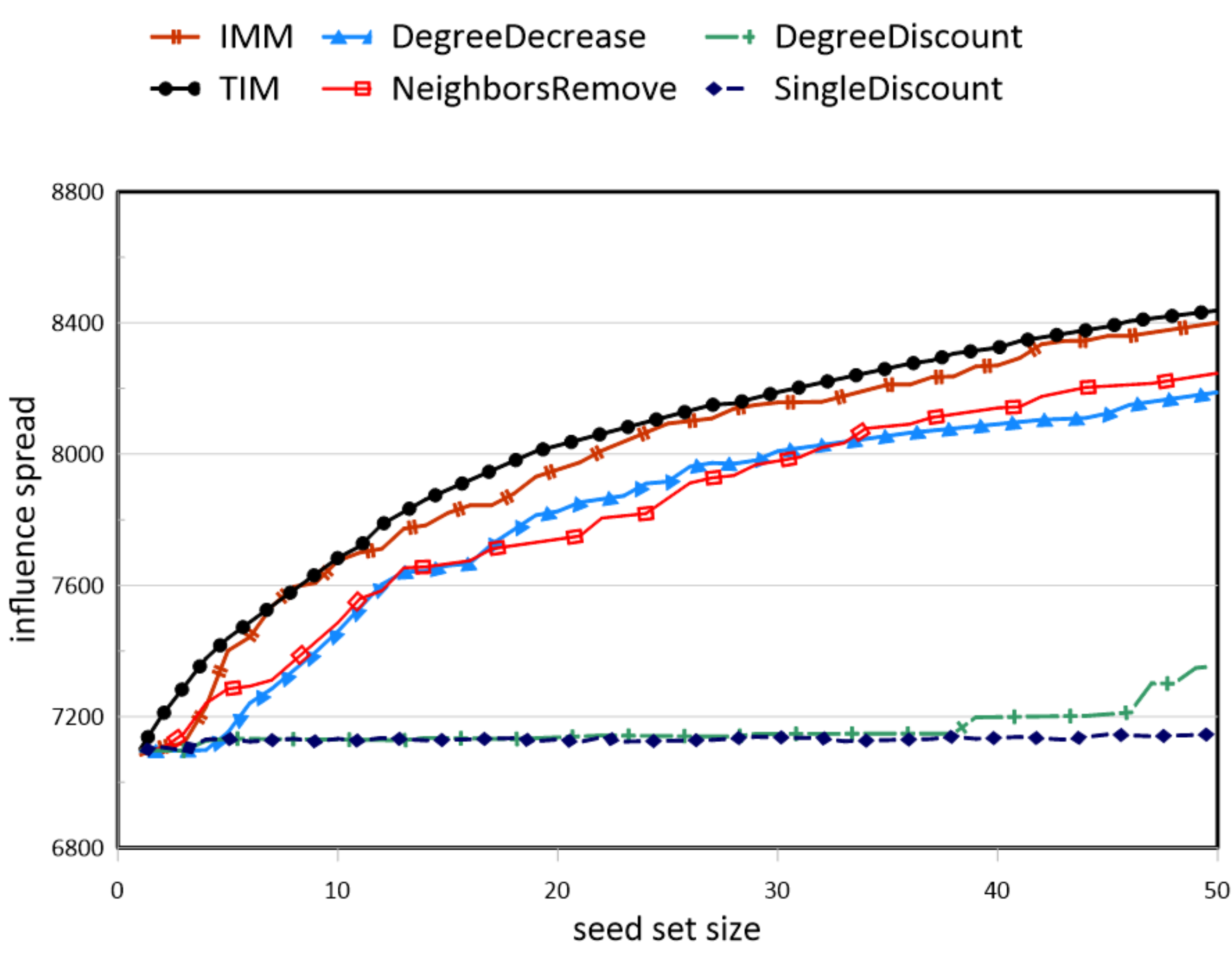}}
		\subcaption{}
		\label{inf_phy_p_10percent}
	\end{subfigure}
	\caption{Running times (a) and influence spreads (b) of algorithms on NetPHY under independent cascade model ($p=0.1$, $k=50$).}
	\label{phy_p_10percent}
\end{figure}

\begin{figure}
	\centering
	\begin{subfigure}[h]{0.49\textwidth}
		\fbox{\includegraphics[width=0.95\linewidth]{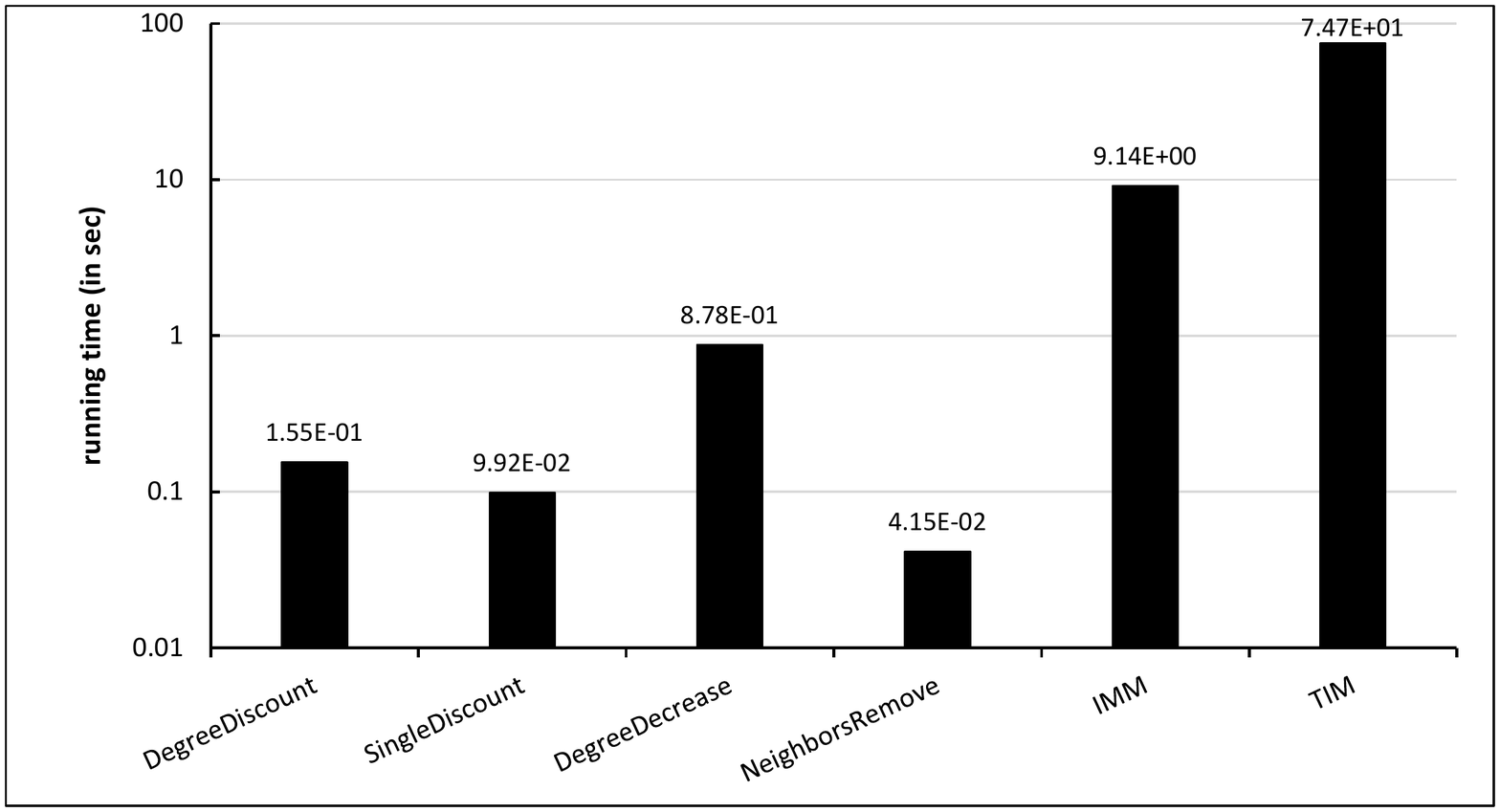}}
		\subcaption{}
		\label{runtime_Epinions_p_1percent}
	\end{subfigure}
	\begin{subfigure}[h]{0.49\textwidth}
		\fbox{\includegraphics[width=.94\linewidth]{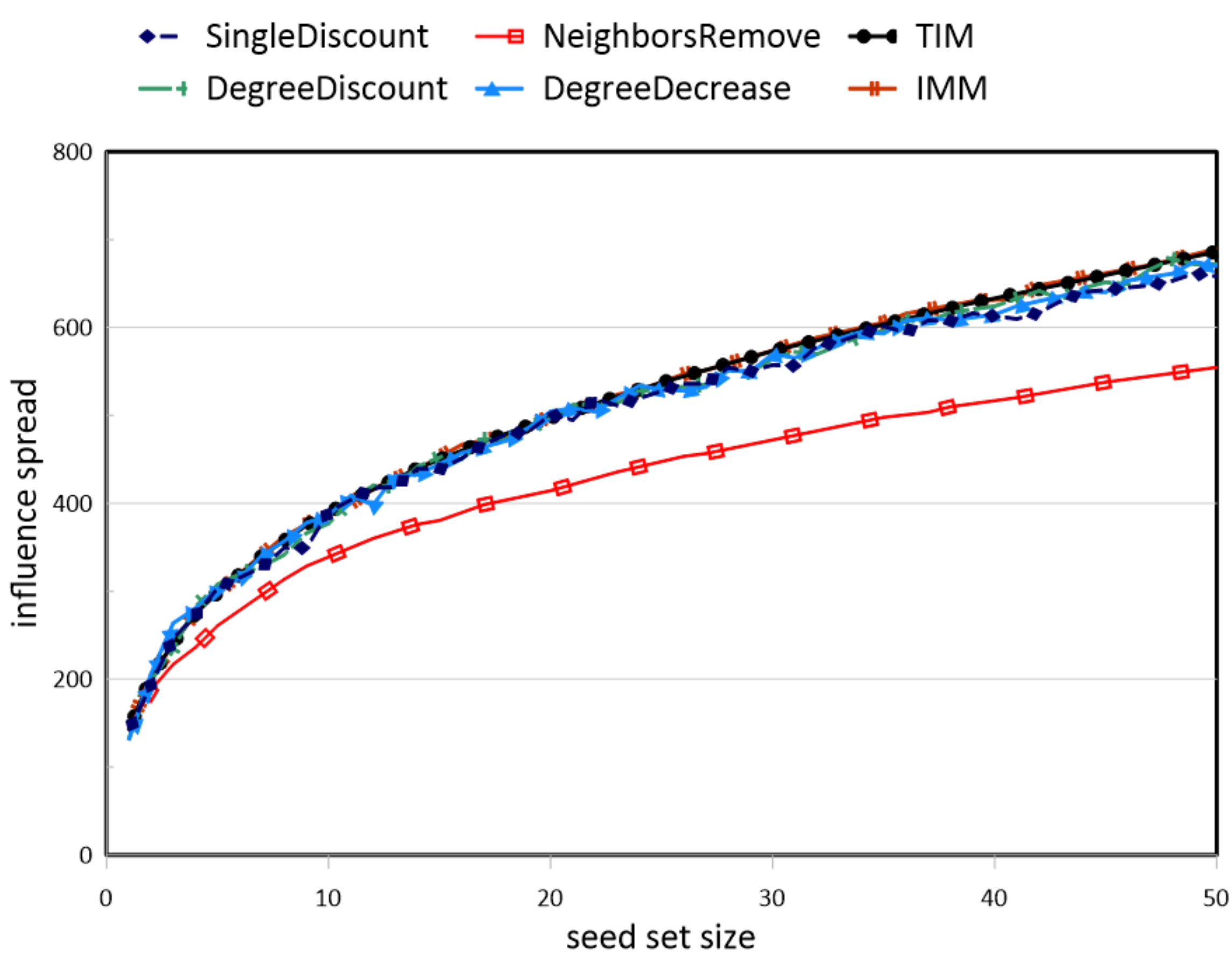}}
		\subcaption{}
		\label{inf_Epinions_p_1percent}
	\end{subfigure}
	\caption{Running times (a) and influence spreads (b) of algorithms on Epinions under independent cascade model ($p=0.01$, $k=50$).}
	\label{Epinions_p_1percent}
\end{figure}

\begin{figure}
	\centering
	\begin{subfigure}[h]{0.49\textwidth}
		\fbox{\includegraphics[width=0.95\linewidth]{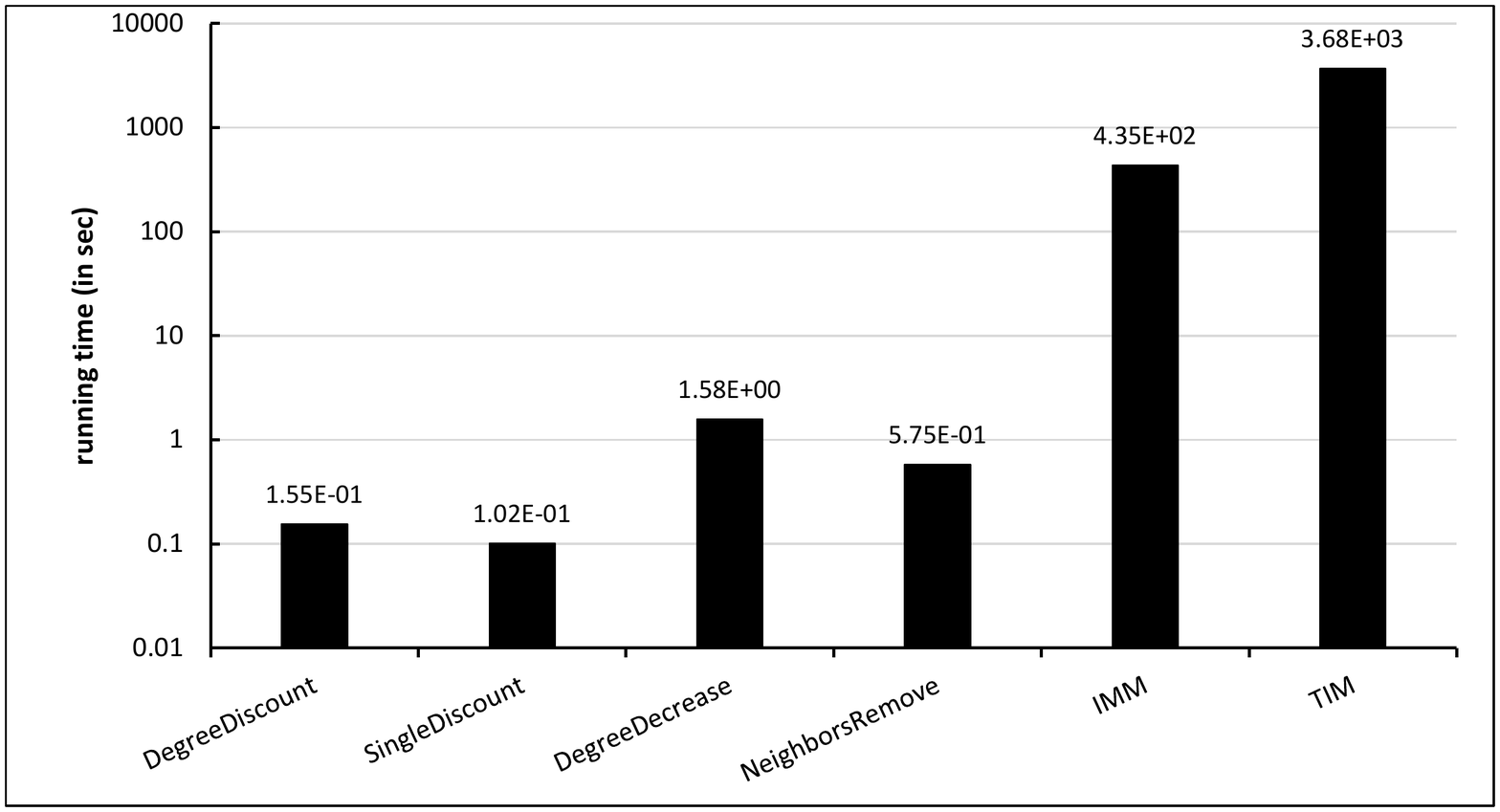}}
		\subcaption{}
		\label{runtime_Epinions_p_10percent}
	\end{subfigure}
	\begin{subfigure}[h]{0.49\textwidth}
		\fbox{\includegraphics[width=.95\linewidth]{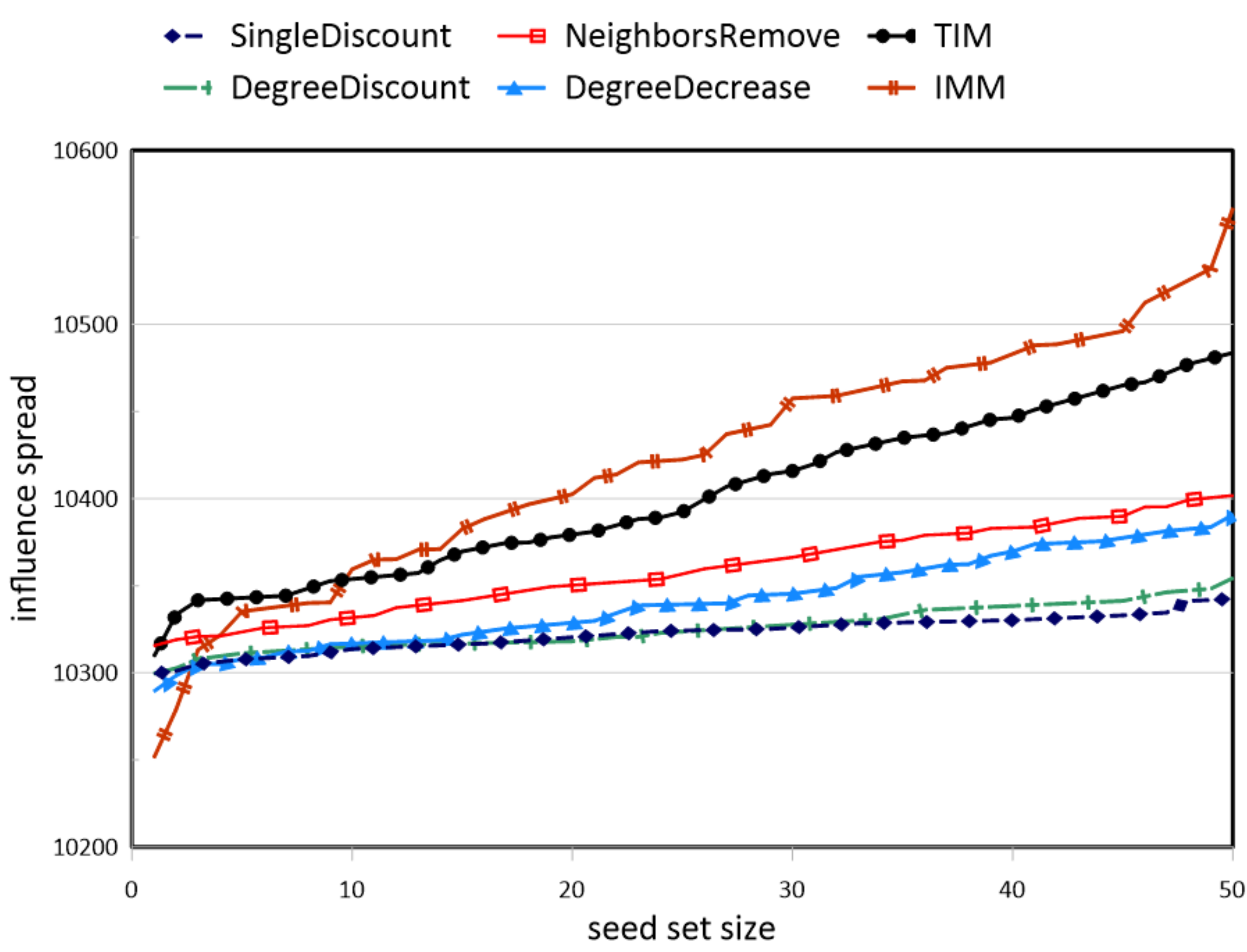}}
		\subcaption{}
		\label{inf_Epinions_p_10percent}
	\end{subfigure}
	\caption{Running times (a) and influence spreads (b) of algorithms on Epinions under independent cascade model ($p=0.1$, $k=50$).}
	\label{Epinions_p_10percent}
\end{figure}

Figure~\ref{hep_p_1percent} and Figure~\ref{hep_p_10percent} show running times and influence spreads of different algorithms under independent cascade model on NetHEPT data-set for $p=0.01$ and $p=0.1$, respectively. Similar results are shown for NetPHY and Epinions data-sets in Figure~\ref{phy_p_1percent}--\ref{Epinions_p_10percent}.

As stated before, we see from the running times charts that the degree centrality heuristics are much faster than TIM and IMM. The running time of DegreeDecrease is usually close to DegreeDiscount and SingleDiscount, while NeighborsRemove is usually faster than all the other algorithms. Sometimes the running time of NeighborsRemove is about 15\% of the running time of the next fastest algorithm.

It can be seen from the influence spread charts, although the proposed algorithms show their superiority for large values of $p$, compared to DegreeDiscount and SingleDiscount heuristics, they still work well even for $p=0.01$ and return solutions of quality close to the quality of solutions of TIM and IMM.

The effectiveness of our algorithms, specially for larger values of $p$ is due to the fact that in those cases the influence of a seed vertex increases on its multi-hop neighbors. Therefore, there will be less advantage from selecting vertices close to the previous seeds. This is exactly one of the main ideas we follow in our proposed algorithms. Our strategy is to avoid selecting vertices with high probability of being influenced.

As it can be seen from the charts, for example Figure~\ref{inf_phy_p_10percent}, the influence propagated by the results of our algorithms is sometimes about 15\% more than the influence propagated by the results of DegreeDiscount and SingleDiscount, while the running times are less than or almost equal to their running times.

\begin{figure}
	\centering
	\fbox{\includegraphics[width=.53\linewidth]{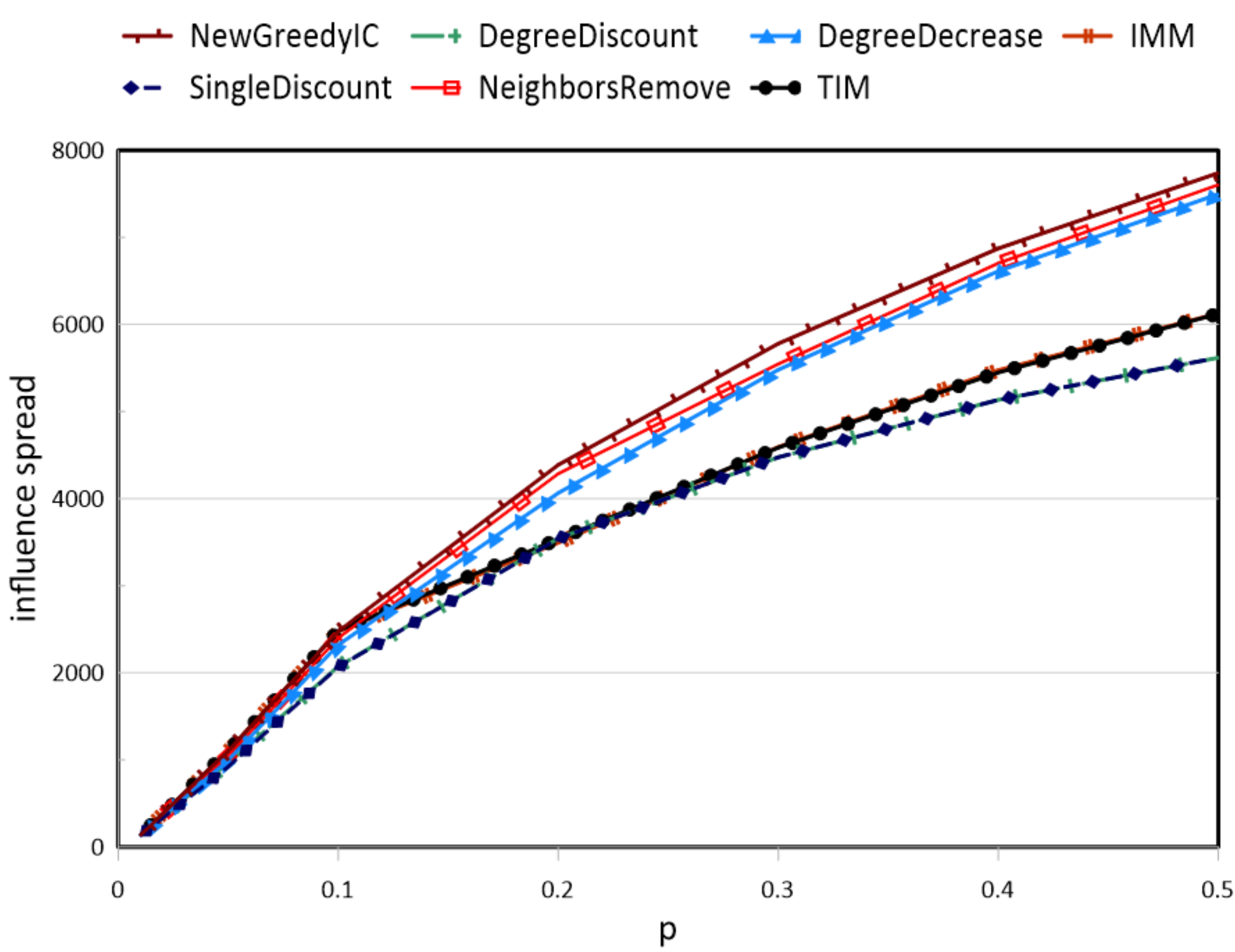}}
	\caption{Comparison of influence spreads of different algorithms on NetHEPT under independent cascade model for different values of $p$}
	\label{inf_p_changes}
\end{figure}

Figure~\ref{inf_p_changes} shows influence spreads of different algorithms under independent cascade model for different values of $p$. As it is seen, from $p=0.12$ on, the influence spread of our algorithms significantly increases, compared to all other algorithms, both degree centrality heuristics and greedy algorithms.

\subsection{Ranking Similarity Analysis}
In this section, we evaluate different influence maximization algorithms in terms of the similarity of the results to the results of the algorithm by Kempe \etal~\cite{kempe2003maximizing}. The similarity between two ranking methods, denoted by $F(k)$, represents the amount of similarity between the results of the methods and is defined as $$F(k)=\frac{L(k)\cap L'(k)}{k},$$ where $L(k)$ and $L'(k)$ are the set of top-$k$ nodes in the two ranking methods.

For Kempe \etal's method~\cite{kempe2003maximizing}, which we use as the true ranking, we consider the result of 20,000 times Monte Carlo simulations and compare the results of other algorithms with this ranking based on the ranking similarity.

\begin{figure}
	\centering
	\begin{subfigure}[h]{0.49\textwidth}
		\fbox{\includegraphics[width=0.95\linewidth]{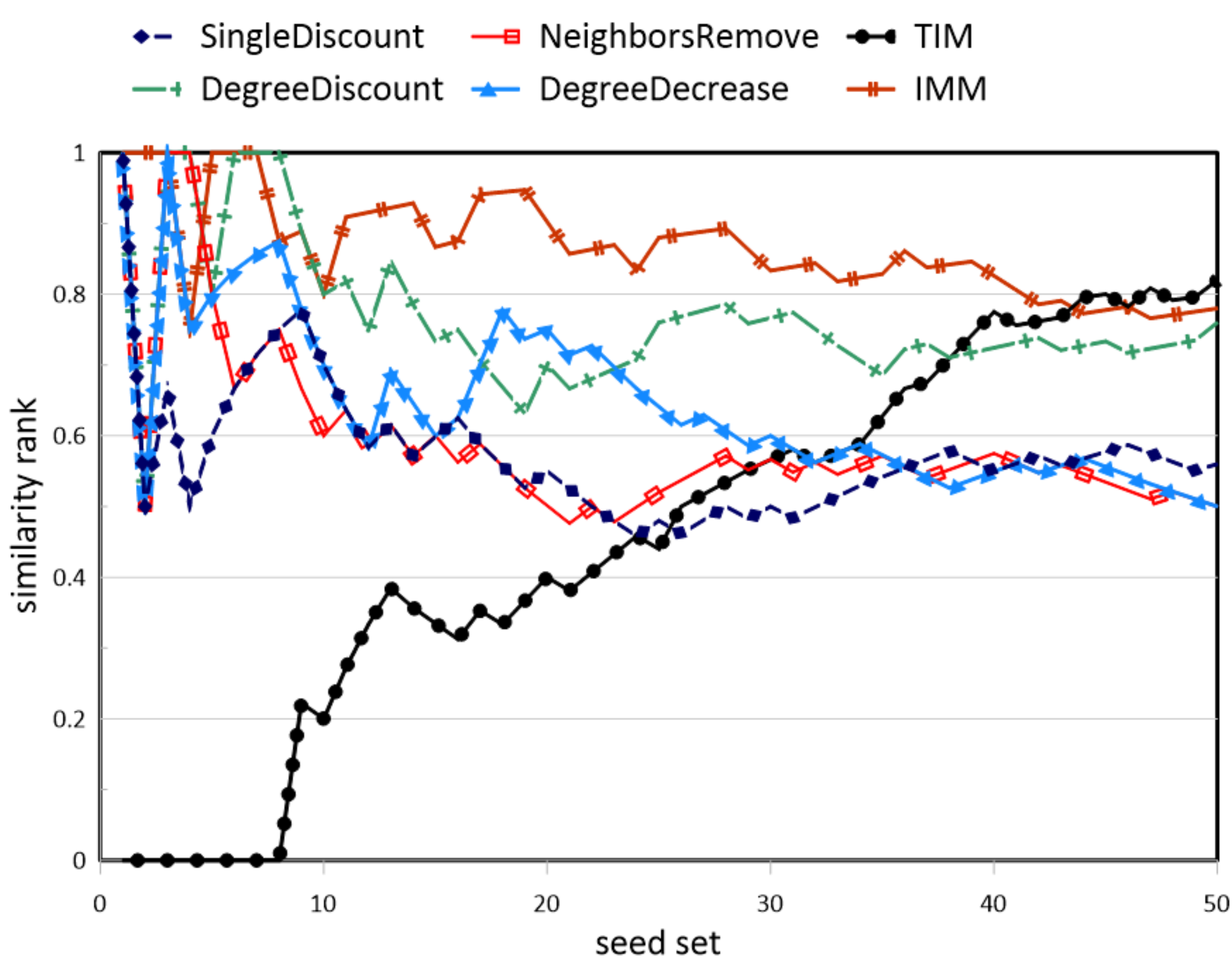}}
		\caption{$p=0.01$}\label{R_hep_0.01}
	\end{subfigure}
	\begin{subfigure}[h]{0.49\textwidth}
		\fbox{\includegraphics[width=.95\linewidth]{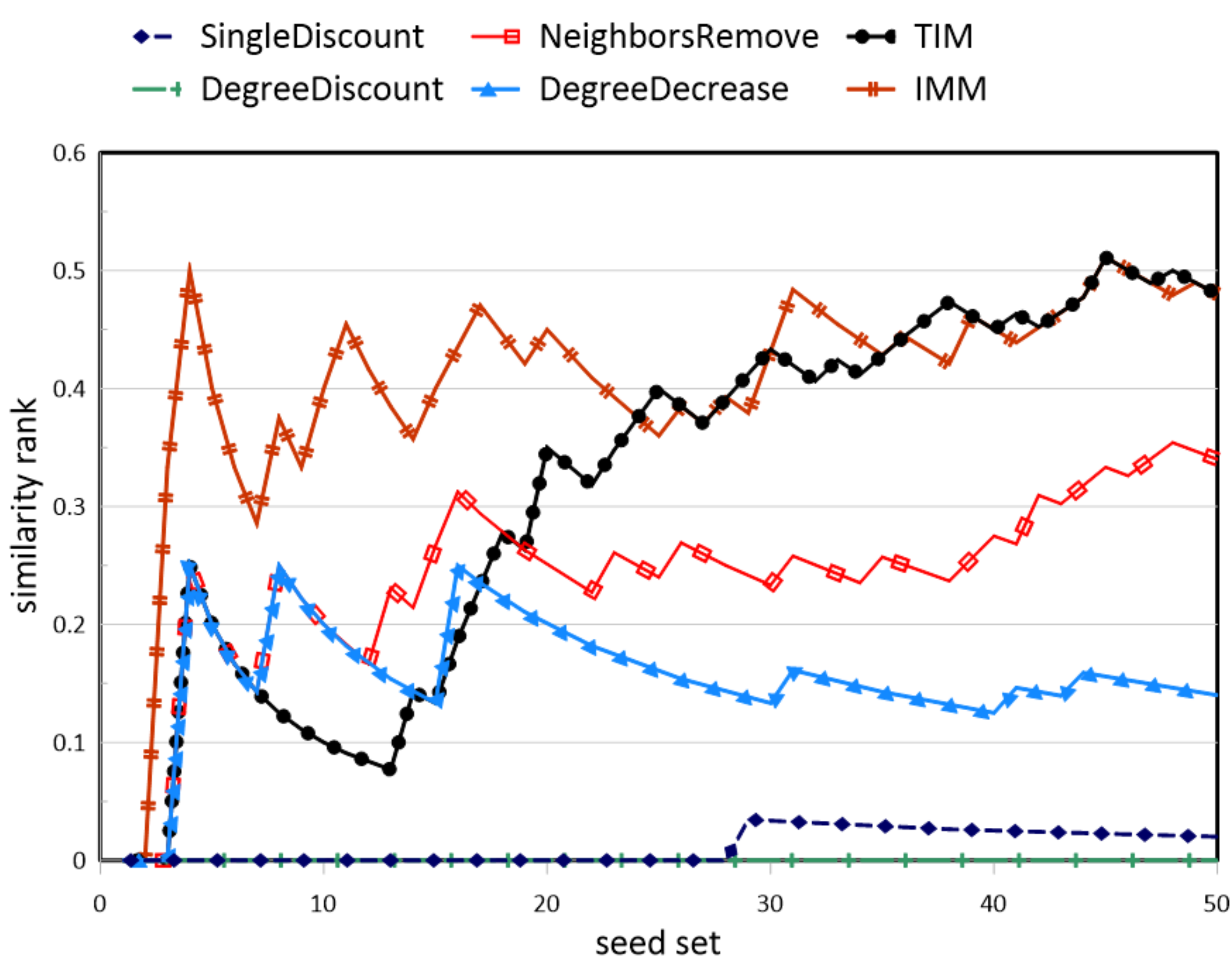}}
		\caption{$p=0.1$}\label{R_hep_0.1}
	\end{subfigure}
	\caption{Rank similarity comparison on NetHEPT.}
	\label{R_hep}
\end{figure}

In Figure~\ref{R_hep}, we see the comparison of ranking similarities on NetHEPT. Figure \ref{R_hep_0.01} shows that for $p=0.01$, IMM and DegreeDiscount have the most ranking similarity to the true ranking. It can also be seen that the results of NeighborsRemove and DegreeDecrease have high similarity to the true ranking in the beginning but as the value of $k$ increases the similarity decreases compared to the other methods. On the other hand, for $p=0.1$, Figure~\ref{R_hep_0.1} shows that DegreeDecrease and especially NeighborsRemove have greater ranking similarity to the true ranking than SingleDiscount and DegreeDiscount, which proves the effectiveness of our methods for larger values of $p$.

\begin{figure}
	\centering
	\begin{subfigure}[h]{0.49\textwidth}
		\fbox{\includegraphics[width=0.95\linewidth]{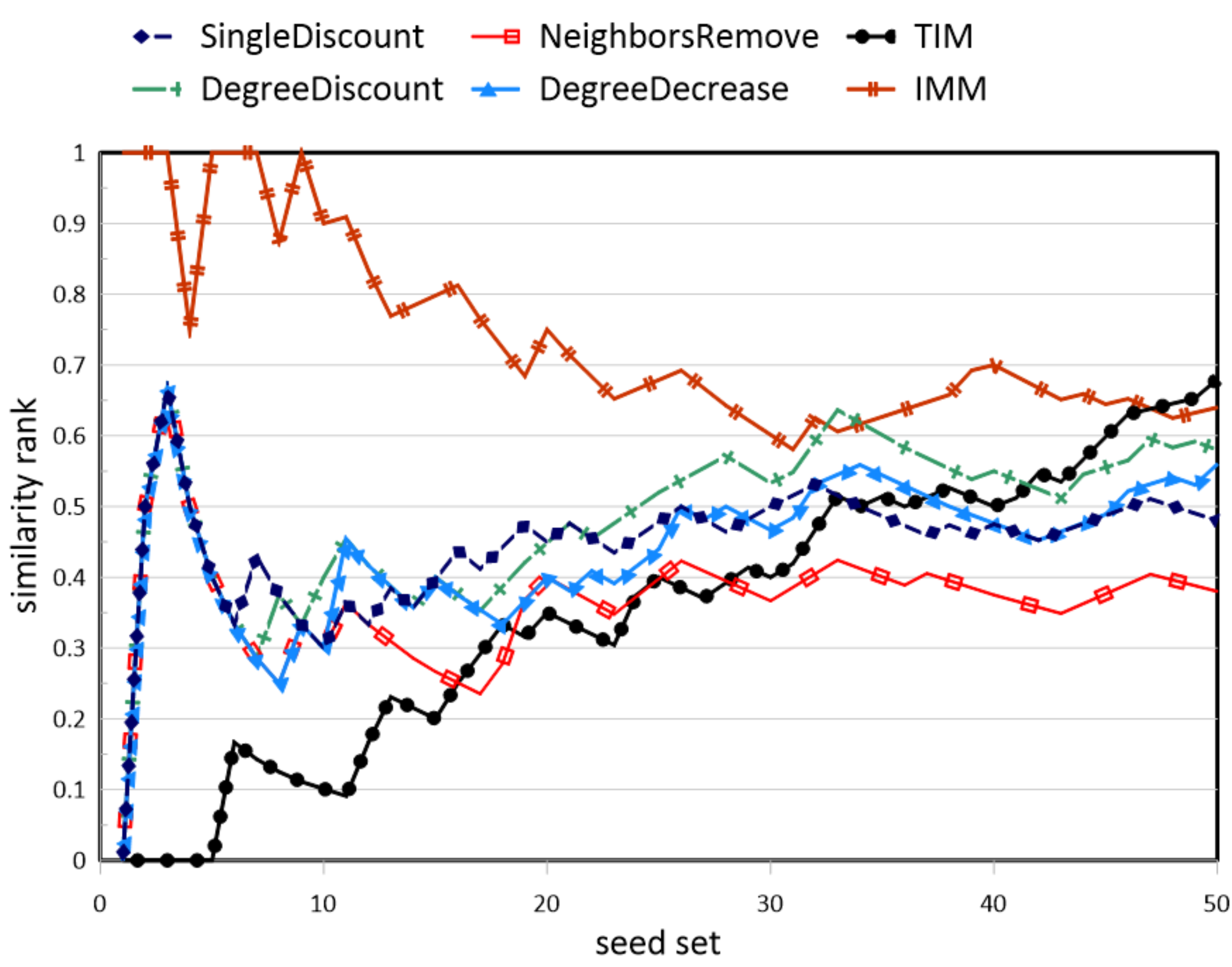}}
		\caption{$p=0.01$}\label{R_phy_0.01}
	\end{subfigure}
	\begin{subfigure}[h]{0.49\textwidth}
		\fbox{\includegraphics[width=.95\linewidth]{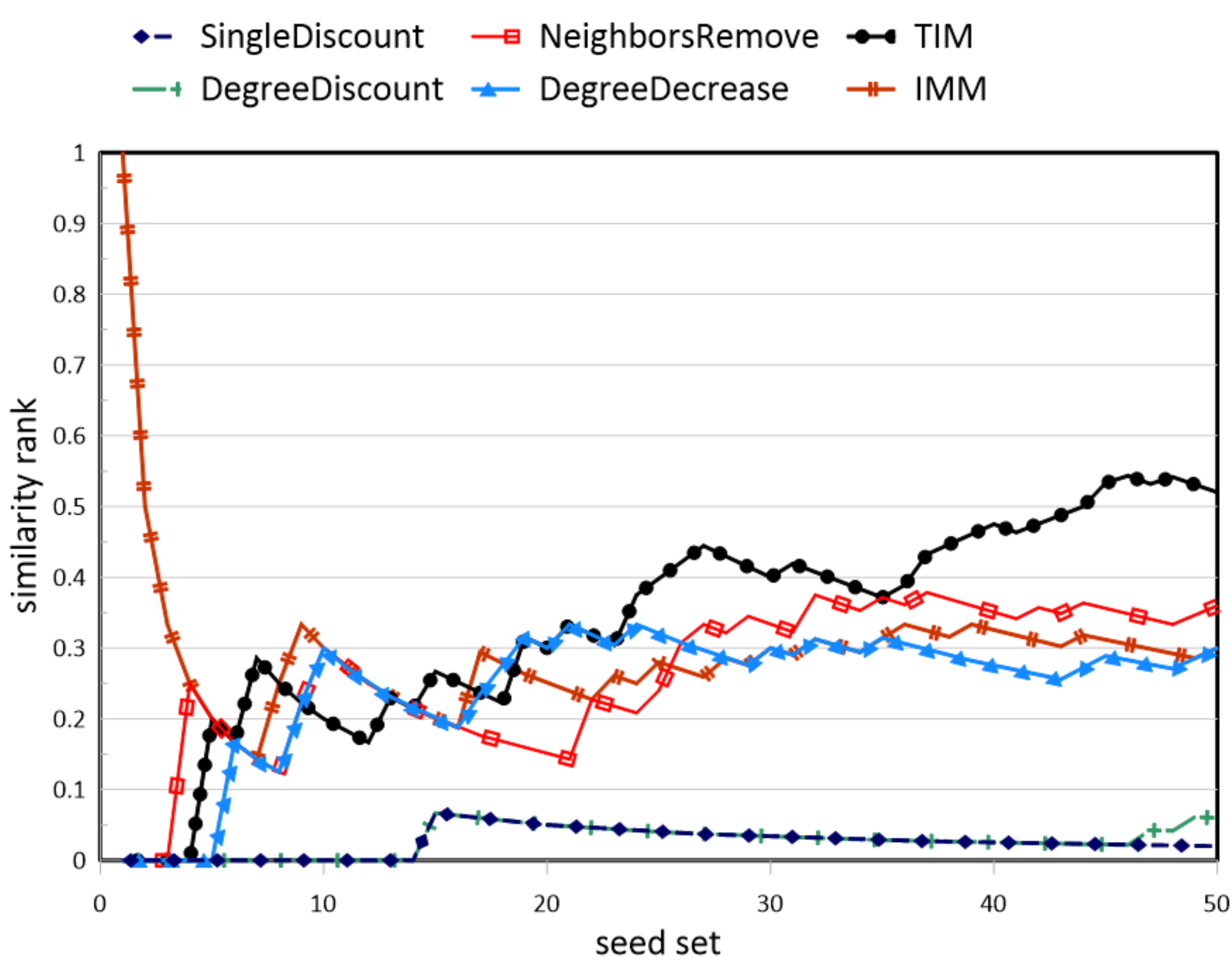}}
		\caption{$p=0.1$}\label{R_phy_0.1}
	\end{subfigure}
	\caption{Rank similarity comparison on NetPHY.}
	\label{R_phy}
\end{figure}

Figure~\ref{R_phy} shows the comparison of ranking similarities on NetPHY. In Figure~\ref{R_phy_0.01} and~\ref{R_phy_0.1}, IMM has the closest ranking to the true one among all methods in the beginning, but as the result size increases, the difference between its ranking and the true ranking tends to increase. However, the other methods show a different behavior. The ranking similarities of all methods are zero at first, and then with the growth in the result size the values tends to increase. As figure~\ref{R_phy_0.1} shows, DegreeDecrease and NeighborsRemove have better ranking in comparison with SingleDiscount and DegreeDiscount for larger values of $p$.

\section{Conclusion} \label{conclusion}
In this paper, we propose two maximum degree based heuristics for influence maximization problem under the independent cascade model. These heuristics take into account the idea that the vertices of high degree are close to each other in social networks. Experiments show that our heuristics outperform previous degree centrality heuristics in terms of the spread of influence in the network.

Since the algorithms that guarantee the quality of the outputs are very time-consuming on large-scale networks, finding heuristics which have small running time and produce solutions of good quality is so desirable. While the influence spread of the results produced by our proposed algorithms are close to the outputs generated by the approximation algorithms, the algorithms run in a much shorter time.

For future work, we will examine the maximum-degree based heuristics for other cascade models. Also, we are looking for more accurate strategies to improve the spread of influence with small running time.

\appendix
\section{Appendix}\label{appendix}

\begin{table}
	\caption{The influence spread of DegreeDecrease for different values of $\alpha$ and $\beta$ on NetPHY for $p=0.01$.}\label{ab_phy_0.01}
	\begin{center}
		\begin{tabular}{|c|c|c|c|c|c|}\hline
			\backslashbox{$\beta$}{$\alpha$}
			& 10 & 20& 30 & 40 & 50\\ \hline
			2 &268.418  &273.005 & 276.168&287.087 & 287.077\\ \hline
			3 &273.309 &277.849 &286.867 &290.409 & 291.459\\ \hline
			5 & 277.808&288.308 & 293.268& 292.866&298.408 \\ \hline
			7 &281.396 &293.107 &291.946 & 300.539& 301.49\\ \hline
			10 &287.748 &294.812 &300.445 & 301.515&305.889 \\ \hline
		\end{tabular}
	\end{center}
\end{table}

\begin{table}\caption{The influence spread of DegreeDecrease for different values of $\alpha$ and $\beta$ on NetPHY for $p=0.01$.}\label{ab_phy_0.1}
	\begin{center}
		\begin{tabular}{|c|c|c|c|c|c|}\hline
			\backslashbox{$\beta$}{$\alpha$}
			& 10 & 20& 30 & 40 & 50\\ \hline
			2 &	7146.63&	7333.55&	7444.47&	7548.64&	7625.16 \\ \hline
			3 &	7341.11	&7615.09&	7737.2&	7789.06&	7840.71\\ \hline
			5 &	7702.44&	8005.31&	8069.25&	8089.31&	8114.76\\ \hline
			7 &	7911.9&	8090.64&	8134.55&	8150.63&	8154.23\\ \hline
			10 &	8065.06&	8100.42&	8159.4&	8177.86&	8189.77\\ \hline
		\end{tabular}
	\end{center}
\end{table}

In this section, we show the results of our experiments to choose the value of parameters of the algorithms. All the experiments are performed to select 50 seeds in the selected network. There are three parameters in DegreeDecrease that are needed to be determined, $\alpha$, $\beta$ and $\epsilon$. We have run several experiments on NetPHY data-sets for $p=0.01$ and $p=0.1$ with different values for $\alpha$ and $\beta$ to find the best combination of values. The influence spreads are shown in Table~\ref{ab_phy_0.01}-\ref{ab_phy_0.1}. Based on the results of the experiments, we find the best selection as $\alpha = 50$ and $\beta = 10$.

The next parameter in DegreeDecrease is $\epsilon$. In Table~\ref{threshold_inf}-\ref{threshold_time} the influence spreads and the running time of DegreeDecrease for different values of $\epsilon$ are reported. From the results, and taking into consideration the fact that selecting a large value for $\epsilon$ may decrease the accuracy of the algorithm on other data-sets, we select the threshold value as $\epsilon=0.1$.

Table~\ref{define_levels} represents the influence spread of NeighborsRemove for different values of $h$ and $p$. The most influence in each row, which has been written in bold, shows the best value for $h$. According to the results we suggest $h = \round{12 \sqrt{p}}$, as stated before.

\begin{table}
	\caption{The influence spread of DegreeDecrease for different values of $\epsilon$.}\label{threshold_inf}
	\begin{center}
		\begin{tabular}{|c|c|c|c|c|c|c|} \hline
			data-set & \multicolumn{2}{c|}{NetHEPT} & \multicolumn{2}{c|}{NetPHY} & \multicolumn{2}{c|}{Epinions} \\ \hline
			\backslashbox{$\epsilon$ ~}{$p$} &0.01&0.1	&0.01&0.1	&0.01&0.1\\ \hline
			&0.01&0.1	&0.01&0.1	&0.01&0.1\\ \hline
			0.01&	127.67&	2322.71&	272.47&	7636.74&	673.76&	10382.9\\ \hline
			0.05&	127.74&	2318.49&	273.19&	7633.28&	672.76&	10382.8\\ \hline
			0.1&	128.34&	2322.83&	275.97&	7638.84&	674.85&	10385.4\\ \hline
			0.5&	128.45&	2320.71&	272.32&	7635.79&	674.19&	10384.2\\ \hline
			1&	127.92&	2321.34&	274.67&	7636.55&	670.25&	10381.4\\ \hline
			3&	127.52&	2320.71&	280.41&	7632.02&	671.55&	10382.1\\ \hline
			5&	129.43&	2322.55&	283.63&	7635.15&	671.91&	10383.6\\ \hline
			7&	128.59&	2322.82&	283.77&	7632.83&	660.47&	10382.7\\ \hline
			10&	128.57&	2319.72&	284.38&	7630.14&	661.04&	10383.6\\ \hline	 	
		\end{tabular}
	\end{center}
\end{table}

\begin{table}
	\caption{The running time of DegreeDecrease for different values of $\epsilon$.}\label{threshold_time}
	\begin{center}
		\begin{tabular}{|c|c|c|c|c|c|c|} \hline
			data-set & \multicolumn{2}{c|}{NetHEPT} & \multicolumn{2}{c|}{NetPHY} & \multicolumn{2}{c|}{Epinions} \\ \hline
			\backslashbox{$\epsilon$ ~}{$p$} &0.01&0.1	&0.01&0.1	&0.01&0.1\\ \hline
			0.01&	0.0735&	0.031&	0.212&	0.090&	0.785&	0.893\\ \hline
			0.05&	0.060&	0.031&	0.156&	0.090&	0.785&	0.889\\ \hline
			0.1&	0.050&	0.031&	0.112&	0.091&	0.498&	0.887\\ \hline
			0.5&	0.043&	0.031&	0.095&	0.091&	0.498&	0.890\\ \hline
			1&	0.041&	0.031&	0.087&	0.092&	0.179&	0.889\\ \hline
			3&	0.034&	0.031&	0.076&	0.090&	0.179&	0.890\\ \hline
			5&	0.031&	0.031&	0.074&	0.091&	0.179&	0.897\\ \hline
			7&	0.031&	0.031&	0.072&	0.091&	0.137&	0.891\\ \hline
			10&	0.031&	0.031&	0.0717&	0.090&	0.137&	0.891\\ \hline
			
		\end{tabular}
	\end{center}
\end{table}

\begin{table}
  \caption{The influence spread of NeighborsRemove for different values of $h$ and $p$.}\label{define_levels}
  \begin{center}
  	\footnotesize
	\begin{tabular}{|c|c|c|c|c|c|c|c|c|c|}\hline
		data-set &\backslashbox{$p$}{$h$}
		& 1 & 2 & 3 & 4 & 5 & 6 & 7 \\ \hline
		\multirow{3}{*}{NetHEPT} & 0.01 & \textbf{127.57}&	101.89&	82.66&	74.80&	72.17&	70.54&	69.7185\\ \cline{2-9}
		&0.05&	981.78&	1056.55&	\textbf{1047.98}&	1012.89&	994.64&	985.25&	975.54\\ \cline{2-9}
		
		& 0.1 &2081.51&	2279.97&	2361.26&	\textbf{2399.12}&	2388.78&	2357.75	&2333.27\\ \hline
		\multirow{3}{*}{NetPHY} & 0.01 &\textbf{280.41}&	219.11&	146.14&	106.47&	97.79&	96.74&	96.14\\ \cline{2-9}
		&0.05&	3832.77&	4155.09&	\textbf{4331.41}&	4288.06&	4213.6&	4210.28&	4204.85\\ \cline{2-9}
		
		& 0.1 &7366.74&	7799.77&	8107.84&	\textbf{8247.58}&	8213.54&	8208.30&	8207.57\\ \hline
		\multirow{3}{*}{Epinions} & 0.01 &\textbf{553.70}&	283.80&	227.55&	226.97&	230.99&	226.901&	227.036\\ \cline{2-9}
		&0.05&	5724.05&	5689.48&	\textbf{5690.16}&	5686.31&	5687.40&	5683.12&	5685.07\\ \cline{2-9}
		
		& 0.1 & 10348.3 &	10426.5 &	10409.4 &	\textbf{10400.9} &	10395.1 &	10399.5	 &10396.4\\ \hline	    				
	\end{tabular}
  \end{center}
\end{table}

\bibliographystyle{plain}
\bibliography{degree}
\end{document}